\documentclass[twocolumn]{aastex631}

\shorttitle{Barred Galaxy Centres}
\shortauthors{Laing et al.}
\graphicspath{{./}{figures/}}

\begin{document}

\title{Increased molecular gas velocity dispersion and star formation efficiency in barred galaxy centres} 

\author{Jennifer M. Laing}
\affiliation{McMaster University
Hamilton, Ontario, Canada}

\author{Christine D. Wilson}
\affiliation{McMaster University
Hamilton, Ontario, Canada}

%



\begin{abstract}

Work by the Physics at High Angular resolution in Nearby GalaxieS (PHANGS) collaboration found higher molecular gas surface densities and velocity dispersions in the centres of barred galaxies compared to unbarred galaxies. We explore central molecular gas using published high resolution (150 pc) measurements of CO$(2-1)$ from the PHANGS-ALMA survey and a new velocity dispersion-dependent prescription for the CO-to-H$_{2}$ conversion factor $\alpha_{\rm{CO}}$. Comparisons of the molecular gas surface density, velocity dispersion, star formation rate, and depletion time reveal that these quantities are different in the centres of barred and unbarred galaxies. Gas depletion times are found to be shorter in barred galaxy centres. Even when we control for the presence of an AGN, the velocity dispersion and depletion time are found to be statistically different between barred and unbarred galaxy centres. The higher velocity dispersion suggests extra non-circular motions, possibly due to the inflow of gas along the bar, that are not constant but must increase as the star formation rate increases.

\end{abstract}





\section{Introduction} \label{sec:intro}

The central region of a star forming galaxy is a dynamic place where physical processes interact on many scales. Some galaxy centres host structures such as a stellar bar or a star forming ring and studies have suggested that these local environmental structures play a role in the evolution of the galaxy \citep{Kormendy2004secularEvol,Maeda2023SFgasrichBars}. Star forming galaxies with a stellar bar are more common in the universe than unbarred galaxies, with over half of spiral galaxies showing a bar in images at optical and near infrared wavelengths (e.g., \citealt{deVaucouleurs1963,Diaz-Garcia2020distSF}). Over the past few decades extensive observation and modeling work has revealed that bars transport gas and stars inwards and in so doing transform the galaxy structure during its evolution \citep{Jogee2004-8Gyr}.

Molecular gas properties on scales of giant molecular clouds (GMCs) where stars form in the cold molecular gas 
\citep{Bolatto2008,Chevance2023} can give insight into how galaxies evolve. Barred galaxies are known to have higher molecular gas surface densities in the central region than unbarred galaxies \citep{Sheth2002CompMolGas}. In addition, molecular gas surface density as well as velocity dispersion have been found to be higher in barred galaxy centres \citep{SunJiayi2020}. We expect to find higher molecular gas surface densities in the centres because the dynamics of the elongated structure drives an inflow of gas along the bar \citep{Sakamoto1999,Jogee2005-Cntr,Diaz-Garcia2021}. This inflow of gas can fuel activity in the central nuclear region such as star formation in a circumnuclear ring or an active galactic nucleus \citep{Sakamoto1999, Combes2019, Garland2024}. We might also expect velocity dispersion to be higher in the centres of barred galaxies due in part to turbulence\footnote{In, this paper, we use the term turbulence to refer to random motions that are  roughly isotropic. Where relevant, we distinguish between the internal velocity dispersion of an individual cloud (internal turbulence) and the cloud-cloud velocity dispersion from random relative motions of individual clouds (population turbulence).} and shear from the extreme environment of the central region.

The star formation rate (SFR) in different regions relates directly to the local physics; for example, star formation can be induced by shocks or stellar feedback or potentially suppressed by inflows of gas \citep{Schinnerer2019}. This relationship between the dynamics of the gas in the interstellar medium and the star formation rate plays a key role in the evolution of galaxies \citep{Krumholz2021}. Whether star formation is observed to be suppressed in bars is still an open question. \cite{Hogarth2024} found lower star formation efficiency (SFE) in barred galaxies with active gas flows along the bar, indicating that the local environment and stage of evolution of the bar plays a role in the star formation. For statistical studies of a large population of barred galaxies, the way that the bar is defined can produce different results. In a recent study, \cite{Maeda2023SFgasrichBars} do a statistical analysis of 18 nearby star forming barred galaxies with a gas-rich, long-bar structure. They separate different regions within the bar (the ends, centre, and along the length of the bar) and find that star formation is suppressed along the length of the bar, while there is more star formation in the ends and central region, as expected \citep{Maeda2023SFgasrichBars}. If the bar is instead defined as one entity that encloses all three parts of the bar, then the gas and star formation measurements can show increases in these properties for the whole bar \citep{Querejeta2021, Diaz-Garcia2021}. This second method of selection often cannot be avoided due to the resolution of the data and distance to galaxies.

The CO-to-H$_{2}$ conversion factor ($\alpha_{\rm{CO}}$) has been shown to depend on the local conditions of the molecular gas, particularly in the extreme environment of galaxy centres \citep{Bolatto2013_KeyCO}. Previous studies of the central region (e.g., \citealt{Querejeta2021,SunJiayi2020,SunJiayi2022multi}) used a metallicity-based $\alpha_{\rm{CO}}$ typically varying as a function of radius. This approach may not capture the physics in galaxy centres, where $\alpha_{\rm{CO}}$ can be suppressed (e.g., \citealt{Sandstrom2013}). Studies of $\alpha_{\rm{CO}}$ variations have found a strong dependence of $\alpha_{\rm{CO}}$ on CO opacity, and in turn the observed linewidth is also found to correlate with the optical depth variations \citep{Israel2020,Teng2023}. In this work we use a new prescription for $\alpha_{\rm{CO}}$ from \cite{Teng2024} which is based on linewidth measurements.

This paper explores the molecular gas in the central region of nearby galaxies, specifically comparing barred and unbarred galaxies. We analyse five properties of the gas and star formation that play an important role in the physics in that region: region-average molecular gas surface density ($\Sigma_{\rm{mol,RA}}$), intensity-weighted mean molecular gas surface density ($\Sigma_{\rm{mol,IW}}$), intensity-weighted mean velocity dispersion ($\sigma_{\rm{v}}$), star formation rate surface density ($\Sigma_{\rm{SFR}}$), and molecular gas depletion time ($t_{\rm{dep}}$). We do a statistical analysis to understand how these quantities vary in the central region between barred and unbarred galaxies. In Section 2 we describe the sample and data products used. In Section 3 we discuss the results and implications of the statistical analysis. Our conclusions are summarised in Section 4.


\begin{deluxetable*}{lccc}
\tablenum{1}
\tablecaption{Mean global galaxy properties\label{tab:GlobalProp}}
\tablewidth{0pt}
\tablehead{
\colhead{} & \colhead{log M$_{*}$/M$_\odot$} & \colhead{log SFR/(M$_\odot$ yr$^{-1}$)} & \colhead{log M$_{\rm{mol}}$/M$_\odot$}
}
\startdata
    Barred & 10.3 $\pm$ 0.06 (0.42) & 0.05 $\pm$ 0.07 (0.47) & 9.12 $\pm$ 0.08 (0.53) \\
	Unbarred & 10.0 $\pm$ 0.09 (0.41) & -0.16 $\pm$ 0.11 (0.53) & 8.92 $\pm$ 0.13 (0.59) \\
\enddata
\tablecomments{Global galaxy properties for 42 barred and 22 unbarred galaxies using data from \citet{SunJiayi2020}; M$_{\rm{mol}}$ uses their metallicity-dependent prescription for $\alpha_{\rm{CO}}$. Mean $\pm$ $\sigma_{\rm{mean}}$ (standard deviation) for each distribution, where $\sigma_{\rm{mean}}=\sigma / \sqrt{N}$. Anderson-Darling tests show no statistically significant differences between the two populations.
}
\end{deluxetable*}

\section{Sample and Data Products} \label{sec:SampleandData}

\subsection{PHANGS-ALMA sample and CO data\label{subsec:data}}

We use published CO($2-1$) data products from the PHANGS-ALMA survey of nearby star forming galaxies at cloud scale resolutions  of 150 pc and pixel scales equivalent to 75 pc \citep{SunJiayi2018,SunJiayi2020,SunJiayi2022multi,Leroy2021b}. Galaxies in the PHANGS-ALMA Large Program were selected to have distances of $d<17$ Mpc, inclinations of $< 75^{\circ}$, global stellar masses in the range $10^{9.27}-10^{11.15}$ M$_{\odot}$, and a specific SFR (sSFR $=$ SFR/M$_{*}$) $> 10^{-11}$ yr$^{-1}$ \citep{Leroy2021b}. All 75 galaxies in the main sample met the criteria above with some uncertainties. The 15 galaxies in the extended sample do not necessarily meet all of the criteria; for example these galaxies generally have low sSFR, lower mass, or high inclination \citep{Leroy2021b}. We selected our galaxies primarily from the main sample, but exclude some irregular galaxies or lower mass galaxies. In total, our sample consists of 64 galaxies, with 61 galaxies from the PHANGS-ALMA Large Program and 3 galaxies (ESO097-013 (Circinus), NGC 253, and NGC 300) from the extended sample \citep{Leroy2021b}. Galaxies in our sample are listed in Table \ref{tab_galaxylist} in Appendix ~\ref{AppendixA}. The barred or unbarred galaxy classification matches \cite{SunJiayi2020} and \cite{Stuber2021}. For barred galaxies we show the percentage of the bar that is included in the central region used in our analysis. Classification of galaxies with an AGN follows \cite{Stuber2021}.

The velocity dispersion of the gas is traced by the width of the CO emission line. We use the ``effective width" from \cite{Leroy2021b}, calculated following the method from \cite{SunJiayi2018} adapted from \cite{Heyer2001ApJ}. The effective width is given by
\begin{equation}
    \sigma_{\rm{v}} = \frac{I_{\rm{CO}}}{\sqrt{2\pi} \ T_{\rm{peak}}},
\end{equation}

\noindent where $T_{\rm{peak}}$ is the measured brightness temperature at the line peak and $I_{\rm{CO}}$ is the CO integrated intensity. The PHANGS-ALMA maps created for $I_{\rm{CO}}$ and $\sigma_{\rm{v}}$ have a high resolution where each pixel in the image represents one sightline and should often contain a single molecular cloud. 

Measurements of velocity dispersion near the centres of galaxies, where the velocity curve rises most steeply, can experience beam smearing, an artificial line broadening caused by the rotational motions of the galaxy. Although the high resolution, cloud-scale observations of PHANGS-ALMA data can somewhat more reliably separate cloud velocity dispersion from beam smearing, the effect can still substantially increase the measured velocity dispersions in the centres. Full kinematic modelling to separate the sources of line broadening is beyond the scope of this work. However, the barred and unbarred samples are drawn from the same mass distribution, which determines beam smearing through the gradient of the rotation curve. As a result, we do not expect beam smearing to affect barred systems differently from unbarred systems.

The molecular gas surface density ($\Sigma_{\rm{mol}}$) is given by \citep{Bolatto2013_KeyCO}: 
\begin{equation}
    \Sigma_{\rm{mol}} = \alpha_{\rm{CO}} \ R^{-1}_{21} \ I_{\rm{CO}} \ \rm{cos} \ \it i,
\end{equation}

\noindent where $\it i$ is the inclination angle and we adopt $R_{21} = 0.65$ as the CO($2-1$)/CO($1-0$) line ratio \citep{Bolatto2013_KeyCO}. We use a linewidth-based conversion factor ($\alpha_{\rm{CO}}$) from \cite{Teng2024}, which is defined as
\begin{equation}
    \rm{log} \alpha_{\rm{CO}} = -0.81 \ \rm{log}(\sigma_{\rm{v}}) + 1.05
\end{equation}

\noindent with $ \sigma_{\rm{v}}$ in units of km s$^{-1}$ and $\alpha_{\rm{CO}}$ in units of $M_{\odot} \ (\rm{K} \ \rm{km \ s^{-1}} \ \rm{pc}^{2})^{-1}$ as expected. Note that for the global molecular gas mass ($M_{\rm{mol}}$) for the galaxies in our sample (Table \ref{tab:GlobalProp}) we use the global CO luminosity ($L_{\rm{CO}}$) and metallicity-dependent $\alpha_{\rm{CO}}$ from \cite{Leroy2021b}, with the value of $R_{21}$ used in the resolved analysis. We briefly consider the effect of using this metallicity-dependent $\alpha_{\rm{CO}}$ on our results for the galaxy centres in Appendix \ref{AppendixB}.

\subsection{Stellar mass and Star formation rate}  \label{subsec:SFandStellarMass}

\cite{Leroy2021b} calculate the stellar mass of these systems/galaxies using near-infrared maps from IRAC 3.6 $\mu$m \citep{Sheth2010} and WISE 3.4 $\mu$m \citep{Leroy2019} in combination with locally estimated mass-to-light ratios. The effective radii (R$_{\rm{eff}}$) are calculated using the derived $\Sigma_{*}$ \citep{Leroy2021b}; the galaxies in the sample have a range in size of 1.1 kpc $< R_{\rm{eff}} <$11.8 kpc.

In this paper we use global star formation rates from \cite{SunJiayi2020}. Resolved SFRs are from \cite{SunJiayi2022multi}, calculated by combining GALEX FUV and WISE 22$\mu$m data to include both exposed and obscured young stars. The global SFR range for the sample is $0.06 - 14.34$ $M_{\odot}$ yr$^{-1}$. 

The global properties of the galaxies in our sample are listed in \cite{SunJiayi2020} and \cite{Leroy2021b}. We show the mean, the uncertainty on the mean, and the standard deviation for several key quantities in Table \ref{tab:GlobalProp}. An Anderson-Darling test indicates that there is no statistically significant difference between the barred and unbarred samples for these global properties.

\subsection{Resolved Central Regions: hexagonal pixels}  \label{subsec:HexagonalPixel}

The structure in the central region of a galaxy can vary dramatically depending on features such as the size of the galactic bulge, whether there is a galactic bar, the strength and size of a galactic bar, or whether the galaxy has an AGN. For example, if the bar is short, it may lie completely inside the central region. In contrast, for a very large bar with a central star forming ring, that ring may dominate the central region.  

The molecular gas surface density and velocity dispersion are measured at the PHANGS-ALMA 150 pc resolution. To analyse these physical properties and compare them with SFR we use the 1.5 kpc hexagonal pixels developed by the PHANGS collaboration. These combine the resolved data together in non-overlapping hexagons which cover the full area of the galaxy. For our analysis, we define the centre of each galaxy to be the area covered by the central hexagonal pixel.\footnote{This is different from \cite{SunJiayi2020} who used the centre masks defined in \cite{Querejeta2021} that are generally smaller than 1.5 kpc.} We find that this region usually contains just the central region of the bar where we would expect to find any star forming rings, but not the full length of the bar where there might be more turbulence as well as higher shear and/or non-circular motions from the inflow of gas along the bar. For galaxies with a central bar, the fraction of the bar contained inside the central hexagonal pixel is listed in Table \ref{tab_galaxylist}. We use the SFR and $\Sigma_{*}$ calculated by \cite{SunJiayi2022multi} 
for each of these pixels. The PHANGS-ALMA data products include $\Sigma_{\rm{mol}}$ and $\sigma_{\rm{v}}$ averages in the hexagonal pixels. The $\sigma_{\rm{v}}$ is calculated as an intensity-weighted mean. Since $\alpha_{\rm{CO}}$ is assumed to be constant inside a given hexagonal pixel, this weighting is equivalent to weighting by surface density. For $\Sigma_{\rm{mol}}$ two averages are provided, an intensity-weighted mean and a simple average and we explore both in our analysis. The $\cos i$ term has been applied to all surface densities to convert them to a face-on projection \citep[see][for details]{SunJiayi2022multi}. Note that we do not apply the completeness correction from \citet{SunJiayi2022multi} to the intensity weighted-mean for $\Sigma_{\rm{mol}}$, as no such correction is available for the other quantities analyzed in this paper.

\begin{figure*}
	\centering
    \includegraphics[height=0.17\textwidth,trim=0cm 0cm 0cm 0cm,clip]{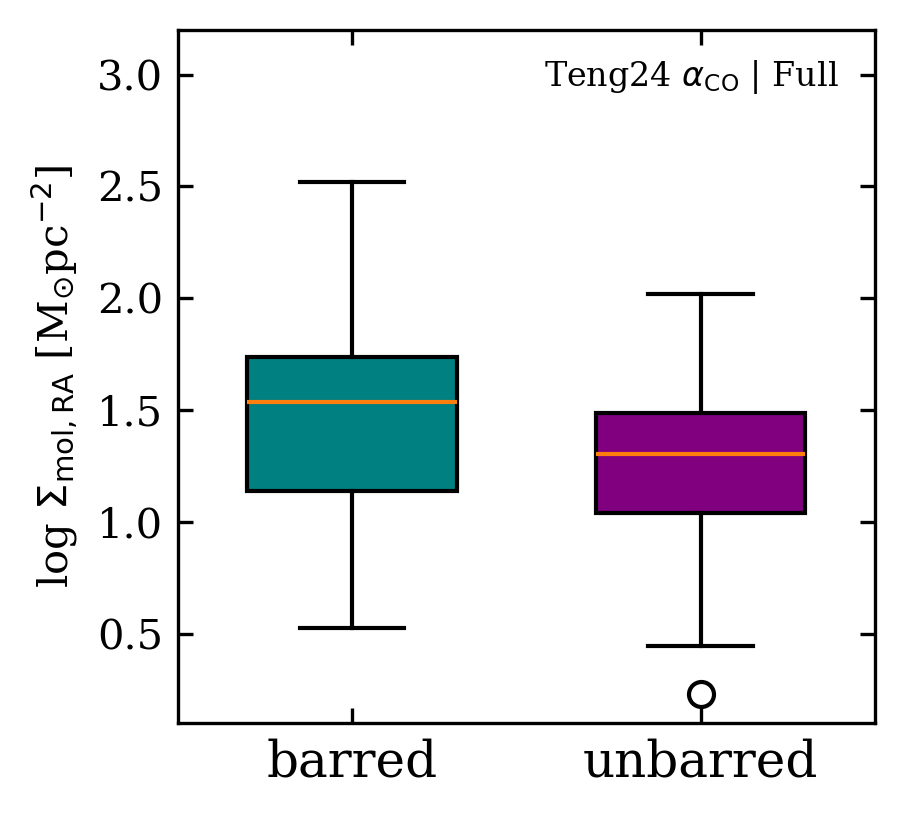} 
    \includegraphics[height=0.17\textwidth,trim=0cm 0cm 0cm 0cm,clip] {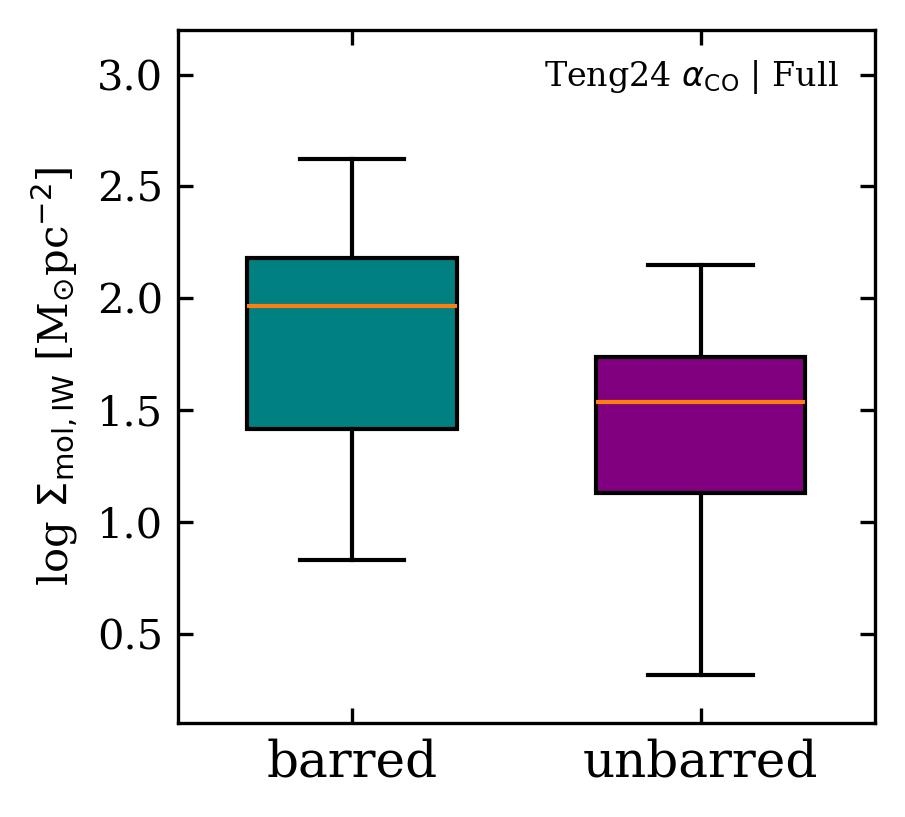} 
    \includegraphics[height=0.17\textwidth,trim=0cm 0cm 0cm 0cm,clip]{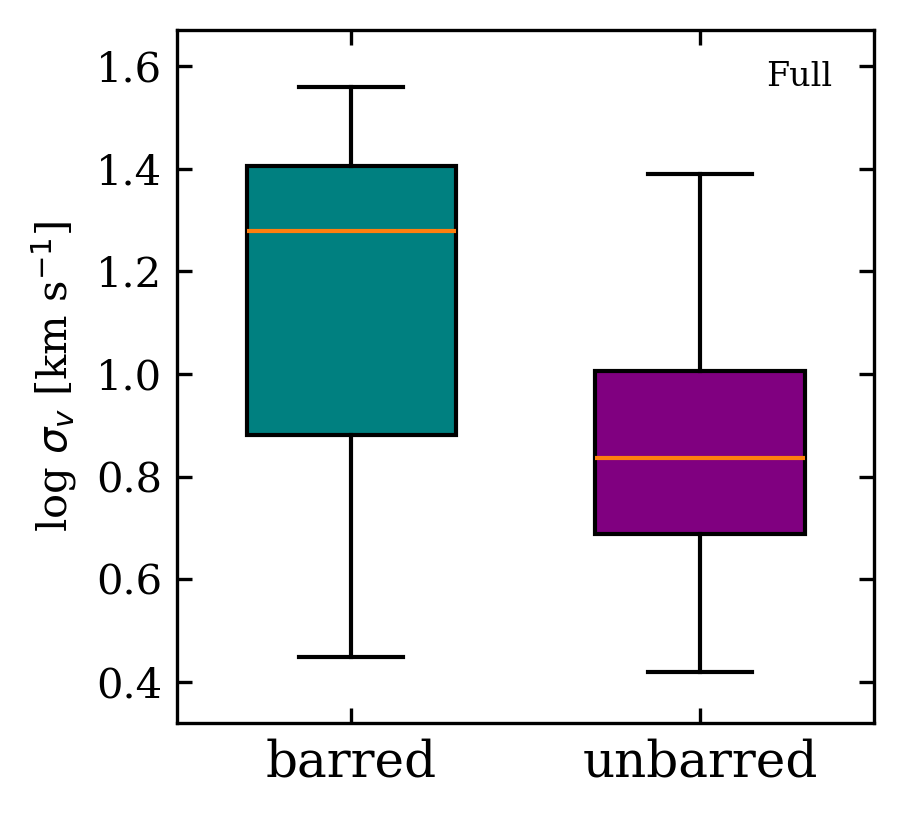}
    \includegraphics[height=0.17\textwidth,trim=0cm 0cm 0cm 0cm,clip]{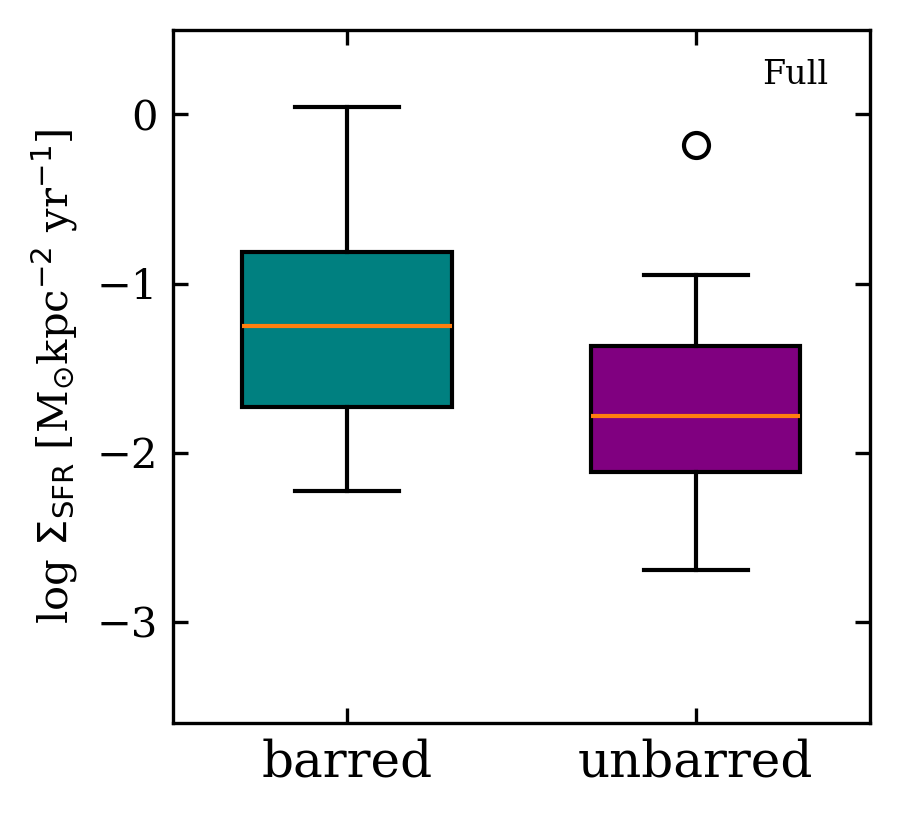}
    \includegraphics[height=0.17\textwidth,trim=0cm 0cm 0cm 0cm,clip]{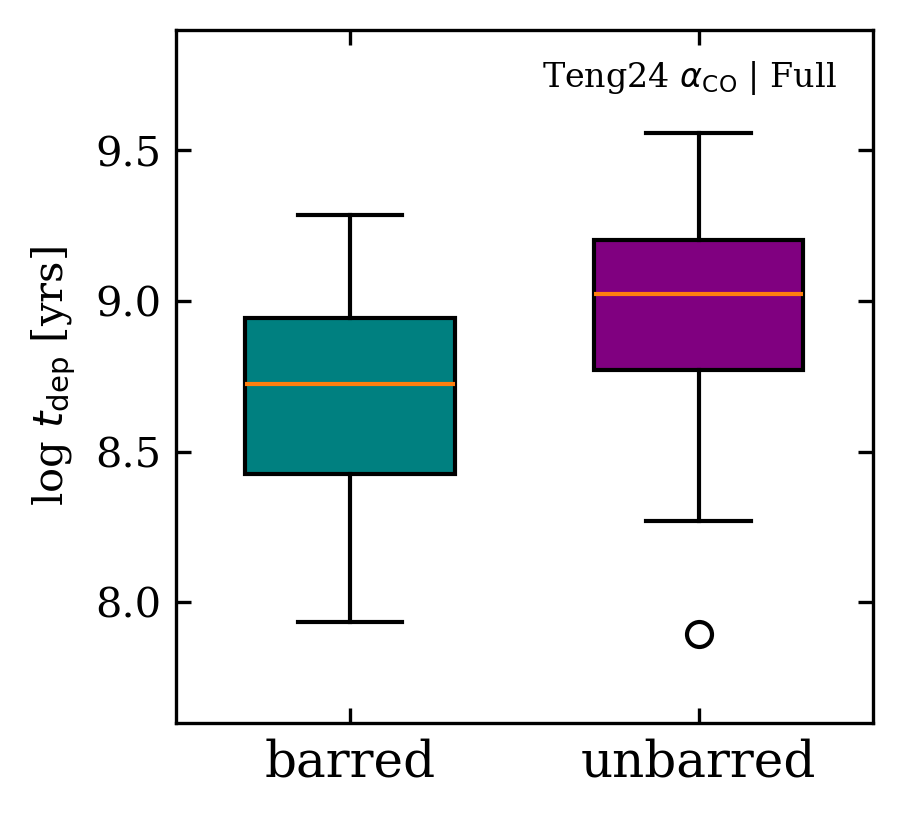} \\
    \includegraphics[height=0.17\textwidth,trim=0cm 0cm 0cm 0cm,clip]{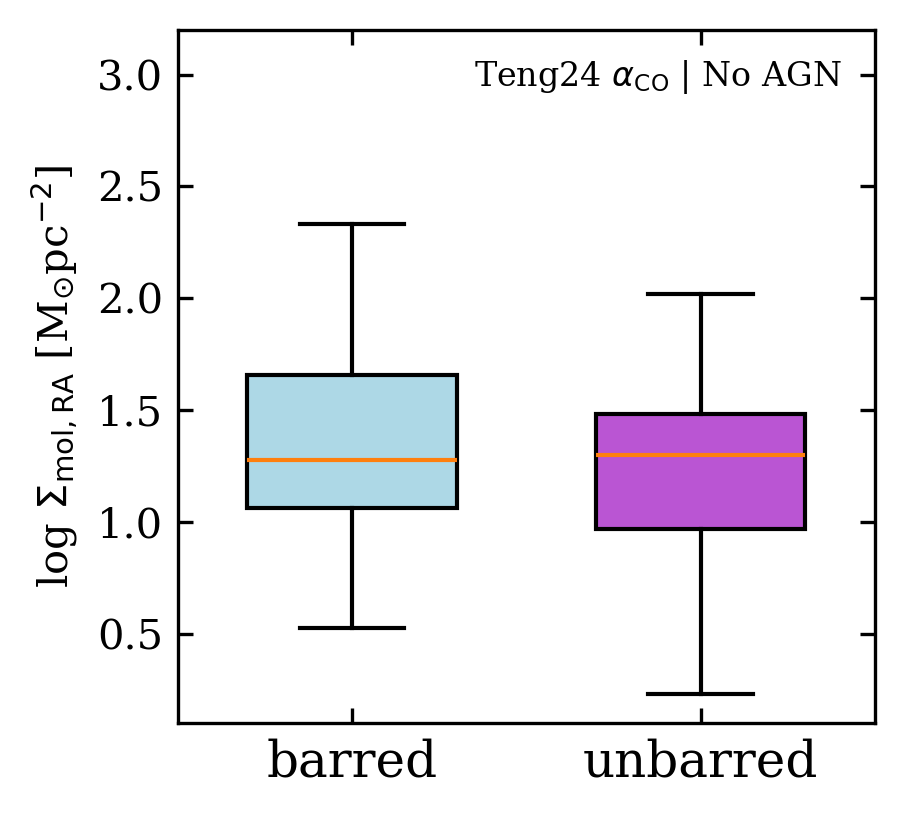} 
    \includegraphics[height=0.17\textwidth,trim=0cm 0cm 0cm 0cm,clip] {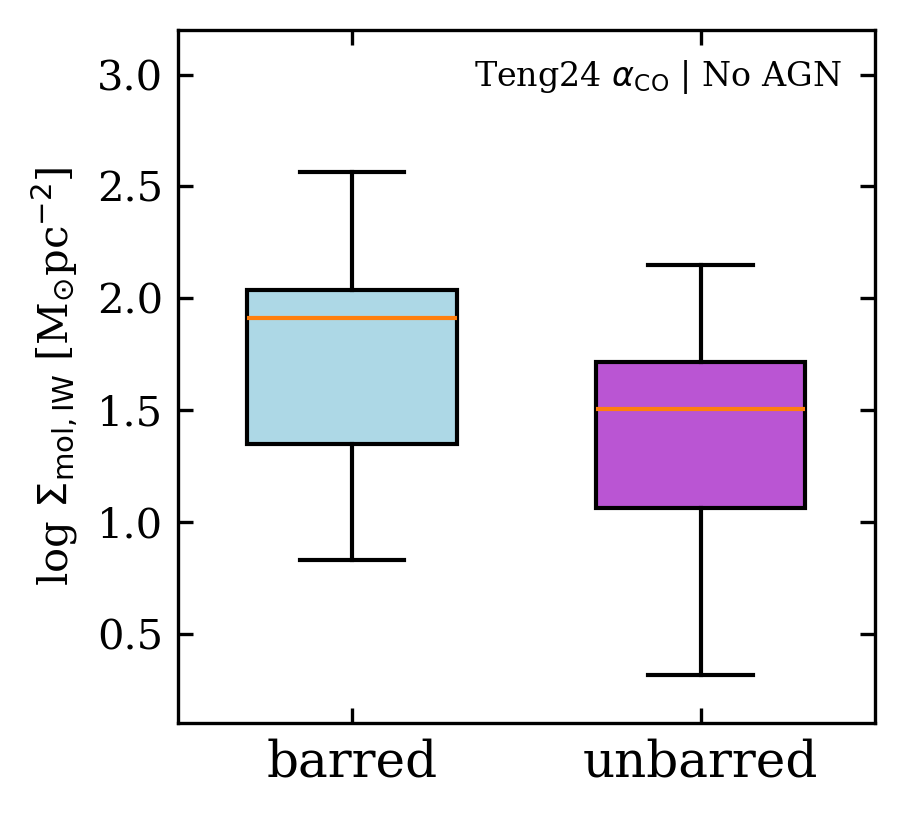}
    \includegraphics[height=0.17\textwidth,trim=0cm 0cm 0cm 0cm,clip]{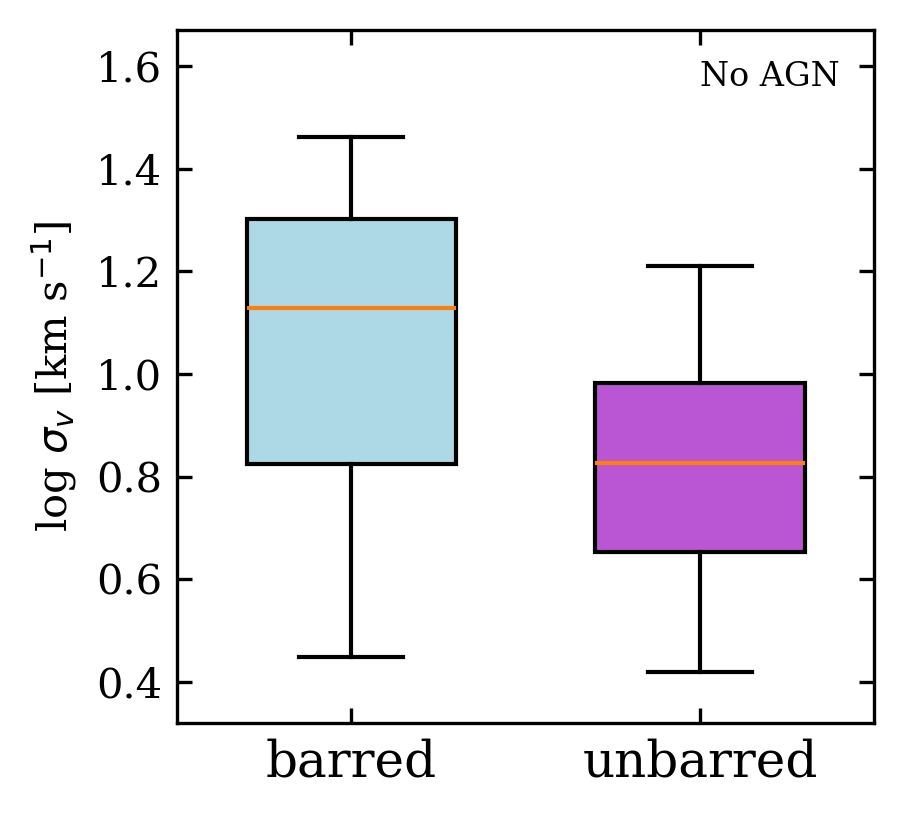}
    \includegraphics[height=0.17\textwidth,trim=0cm 0cm 0cm 0cm,clip]{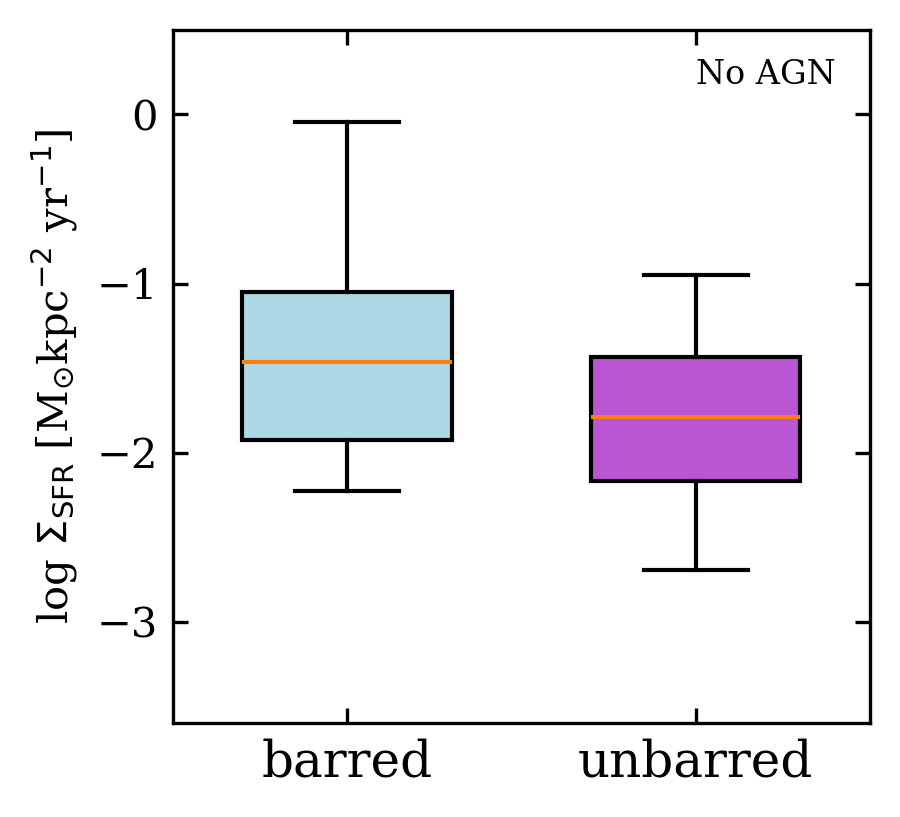}
    \includegraphics[height=0.17\textwidth,trim=0cm 0cm 0cm 0cm,clip]{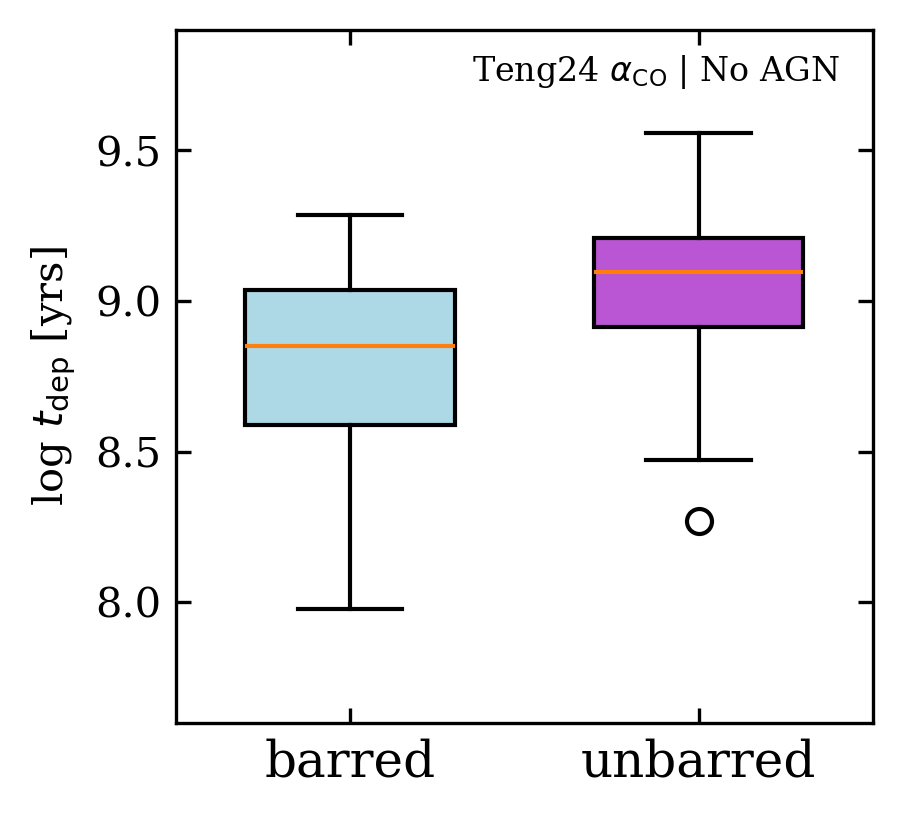}
    \caption{Boxplots comparing the distributions of properties of the central regions for the full sample of barred (42) and unbarred (22) galaxies for each property [top row] and for a subset of the sample containing only galaxies with no AGN, leaving barred (30) and unbarred (19) [bottom row]. From left to right, the plots show: area-averaged molecular gas surface density; intensity-weighted molecular gas surface density; intensity-weighted velocity dispersion; SFR surface density; and gas depletion time (both area-averaged). The box encloses the region between the first and third quartiles with the median line shown in orange. Whiskers extend out to 1.5 times the interquartile range and black circles indicate outliers. The gas calculations are done using the linewidth-based $\alpha_{\rm{CO}}$ from \cite{Teng2024}. We find moderate to strong differences between the two full samples for $\Sigma_{\rm{mol,IW}}$, $\sigma_{\rm{v}}$, $\Sigma_{\rm{SFR}}$, and $t_{\rm{dep}}$. However, when galaxies with an AGN are removed from the sample, the only significant differences between the two distributions are found for $\sigma_{\rm{v}}$ and $t_{\rm{dep}}$.  }
\label{fig:boxResolvedT24}
\end{figure*}

\begin{deluxetable*}{lcccc}
\tablenum{2}
\tablecaption{Resolved galaxy properties: central region  \label{tab:ResolvedPropT24}}
\tablewidth{0pt}
\tablehead{
\colhead{} & \multicolumn{2}{c}{Full Sample} & \multicolumn{2}{c}{Excluding AGN} \\
\colhead{} & \colhead{Barred} & \colhead{Unbarred} & \colhead{Barred} & \colhead{Unbarred}
}

\startdata
	log $\Sigma_{\rm{mol,RA}}$ [M$_{\odot}$ pc$^{-2}$] & 
    1.45$\pm$0.07 (0.44) & 1.26$\pm$0.09 (0.43) & 1.37$\pm$0.08 (0.42) & 1.23$\pm$0.10 (0.45) \\
	log $\Sigma_{\rm{mol,IW}}$ [M$_{\odot}$ pc$^{-2}$] & 
    1.82$\pm$0.07 (0.48) & 1.45$\pm$0.10 (0.47) & 1.70$\pm$0.08 (0.45) & 1.41$\pm$0.10 (0.47) \\
    log $\sigma_{\rm{v}}$ [km s$^{-1}$] & 
    1.17$\pm$0.05 (0.30) & 0.84$\pm$0.06 (0.27) & 1.06$\pm$0.05 (0.30) & 0.79$\pm$0.05 (0.23) \\
    log $\Sigma_{\rm{SFR}}$ [M$_{\odot}$ yr$^{-1}$ kpc$^{-2}$] & 
    -1.24$\pm$0.10 (0.61) & -1.70$\pm$0.13 (0.59) & -1.42$\pm$0.10 (0.54) & -1.79$\pm$0.12 (0.52) \\
    log $t_{\rm{dep}}$ [yr] & 
    8.69$\pm$0.06 (0.36) & 8.96$\pm$0.08 (0.37) & 8.79$\pm$0.06 (0.31) & 9.02$\pm$0.07 (0.31) \\
\enddata
\tablecomments{Resolved gas properties in the central region of the 42 barred and 22 unbarred galaxies in the full sample, and a subset of the sample containing only galaxies with no AGN, leaving 30 barred and 19 unbarred galaxies. Mean $\pm$ $\sigma_{\rm{mean}}$ (standard deviation) for each distribution, where $\sigma_{\rm{mean}}=\sigma / \sqrt{N}$. See Figure \ref{fig:boxResolvedT24} for description of quantities; gas calculations are done using the linewidth-based $\alpha_{\rm{CO}}$ from \cite{Teng2024}. We find significant differences between barred and unbarred galaxy centres for all quantities except $\Sigma_{\rm{mol,RA}}$ for the full sample, and only for $\sigma_{\rm{v}}$ and $t_{\rm{dep}}$ when AGN galaxies are removed (see Table \ref{tab:ADbothSamplesT24}).}
\end{deluxetable*}


\section{Statistical Evidence for Differences between barred and unbarred galaxy centres} \label{sec:StatsandResults}

\subsection{Properties of the Central Regions}  \label{subsec:PropCentres}

\subsubsection{Full Sample}

The five key properties explored in our analysis are region-averaged $\Sigma_{\rm{mol,RA}}$, intensity-weighted mean $\Sigma_{\rm{mol,IW}}$, intensity-weighted mean $\sigma_{\rm{v}}$, region-averaged $\Sigma_{\rm{SFR}}$, and molecular gas depletion time ($t_{\rm{dep}}=\Sigma_{\rm{mol,RA}}/\Sigma_{\rm{SFR}}$) for the central hexagonal pixel. Since the barred and unbarred galaxy samples consist of 42 and 22 galaxies, respectively, we compare the two samples using the Anderson-Darling test (AD-test) for continuous distributions, which is well suited for small samples sizes of less than 50 members \citep{Scholz1987Stats}. The null hypothesis for the AD-test is that the samples are drawn from the same distribution. We accept the null hypothesis for $p$-values $>0.05$. For $p$-values between 0.01 and 0.05 we have moderate evidence against the null hypothesis, and for $p$-values $<0.01$ we have strong evidence against the null hypothesis, suggesting that the distributions are not drawn from the same population. We use the package kSamples in R Studio to run the AD-test for multiple distributions which returns the AD criterion for k samples (where k=2) as well as the AD-test statistic and $p$-value\footnote{https://rdrr.io/cran/kSamples/man/ad.test.html}.

\begin{deluxetable}{lcccc}
\tablenum{3}
\tablecaption{Comparing barred and unbarred galaxy centres: 
full sample and sample which excludes galaxies with an AGN \label{tab:ADbothSamplesT24}}
\tablewidth{0pt}
\tablehead{
\colhead{} & \multicolumn{2}{c}{Full Sample} & \multicolumn{2}{c}{No AGN} \\
\multicolumn{1}{l}{Property} & \colhead{AD stat} & \colhead{$p$-value} & \colhead{AD stat} & \colhead{$p$-value} 
} 
\startdata
$\Sigma_{\rm{mol,RA}}$ & 0.47 & 0.21 & -0.46 & 0.60  \\ 
$\Sigma_{\rm{mol,IW}}$ & 4.0 & 0.0086 & 1.4 & 0.081  \\
$\sigma_{\rm{v}}$ & 8.3 & 0.00023 & 5.4 & 0.0025  \\
$\Sigma_{\rm{SFR}}$ & 3.5 & 0.013 & 1.8 & 0.058  \\ 
$\Sigma_{*}$ & 1.7 & 0.065 & 0.38 & 0.24  \\
$t_{\rm{dep}}$ & 4.7 & 0.0046 & 2.9 & 0.021   \\
$\sigma_{\rm{v}}$/$\Sigma_{\rm{mol,IW}}$ & $-0.9$ & 0.95  & -0.73 & 0.81  \\ 
$\sigma_{\rm{v}}$/$\Sigma_{\rm{SFR}}$ & 0.12 & 0.31 & -0.49 & 0.62 \\
\enddata
\tablecomments{Statistical results of the Anderson-Darling test comparing 42 barred, 22 unbarred galaxies (full sample) and 30 barred, 19 unbarred galaxies with no AGN for each property. Gas calculations are done using the linewidth-based $\alpha_{\rm{CO}}$ from \cite{Teng2024}. We show the AD statistic and $p$-value. For the full sample, quantities with $p$-values $<0.05$ ($\Sigma_{\rm{mol,IW}}$, $\sigma_{\rm{v}}$, $\Sigma_{\rm{SFR}}$ and $t_{\rm dep}$) have evidence against the null hypothesis, which implies that barred and unbarred galaxy centres are likely significantly different populations. When galaxies with an AGN are removed from the sample $\sigma_{\rm{v}}$ and $t_{\rm dep}$ still show evidence that the barred and unbarred populations are different.}
\end{deluxetable}

Figure \ref{fig:boxResolvedT24} shows each of the five properties in boxplots comparing the barred and unbarred galaxy distributions (full sample in the top row and a subset of the sample excluding galaxies with an AGN in the bottom row). The mean, uncertainty on the mean and standard deviation for each distribution are listed in Table \ref{tab:ResolvedPropT24}. For the full sample, we see a significantly higher median value (orange line) for barred galaxies compared to unbarred galaxies for the intensity-weighted $\Sigma_{\rm{mol,IW}}$ (the second boxplot) while for the region averaged $\Sigma_{\rm{mol,RA}}$ (first boxplot) there is no difference between the distributions. This difference between the barred and unbarred distributions is even more evident in the third boxplot showing the intensity-weighted $\sigma_{\rm{v}}$. The $t_{\rm dep}$ is also significantly higher for barred versus unbarred galaxies, while the difference is a bit less for $\Sigma_{\rm{SFR}}$. 

These trends are confirmed in Table \ref{tab:ADbothSamplesT24} which shows the results of the AD-tests comparing barred and unbarred galaxies. For the full sample, four properties ($\Sigma_{\rm{mol,IW}}$, $\sigma_{\rm{v}}$, $\Sigma_{\rm{SFR}}$, and $t_{\rm dep}$) show moderate to strong evidence against the null hypothesis suggesting that they are not from the same population. On the other hand, the AD-test for $\Sigma_{\rm{mol,RA}}$ (as well as for the stellar mass surface density ($\Sigma_{*}$) which we ran for comparison) suggests that the central pixels of the barred and unbarred populations are the same. Two of the properties that show the strongest evidence are the ones that are intensity-weighted over the hexagonal pixel. This is noteworthy because the two intensity-weighted means require the availability of high resolution measurements, while the other quantities are (or, in the case of $\Sigma_{\rm{mol,RA}}$, could have been) derived from lower-resolution data. Thus, the availability of higher resolution data is important for 
the statistics.

We also ran statistical tests on the ratios of the other quantities investigated in this analysis, namely $\sigma_{\rm{v}}/\Sigma_{\rm{mol,IW}}$ and $\sigma_{\rm{v}}/\Sigma_{\rm{SFR}}$. This compares the properties $\sigma_{\rm{v}}$ and $\Sigma_{\rm{mol,IW}}$, which were both higher in barred galaxies, as well as the less well studied relation between $\sigma_{\rm{v}}$ and $\Sigma_{\rm{SFR}}$. The AD-test results show that the barred and unbarred populations are statistically the same for both ratios. 

\subsubsection{Removing Galaxies with an AGN}

The presence of an active galactic nucleus (AGN) could produce increased CO intensity and inferred molecular gas surface density \citep{Bolatto2021,Liu2023}. Studies using hydrodynamical simulations as well as observations show that galactic outflows caused by AGN can reach high speeds ($\geq 1000$ km s$^{-1}$) and extend from the centre of the galaxy to scales in the range of $1-10$ kpc \citep{Costa2014}. This would increase the measured velocity dispersion in the CO observations, while shock-driven compression of the gas caused by outflows can induce increased star formation \citep{Stuber2021}. To see if AGN are the main cause of the differences between the barred and unbarred galaxy populations, we repeat our statistical analysis with a new sub-sample with all of the galaxies with an AGN removed. AGN are identified using optical spectroscopy \citep{Veron-Cetty2010}, and we use the classification of which galaxies have an AGN from \cite{Stuber2021} (Table \ref{tab_galaxylist}). The new sample with AGN removed contains 30 barred galaxies and 19 unbarred galaxies. 
 
Figure \ref{fig:boxResolvedT24} shows boxplots of the distributions of barred and unbarred galaxies without AGN and Tables \ref{tab:ResolvedPropT24} and \ref{tab:ADbothSamplesT24} show the means and the results of the AD-test for the sample of barred versus unbarred galaxies without AGN. When AGN are removed $\sigma_{\rm{v}}$ and $t_{\rm{dep}}$ continue to be different.  $\Sigma_{\rm{mol,IW}}$ and $\Sigma_{\rm{SFR}}$ now show no significant differences. Comparing these with the boxplots in Figure \ref{fig:boxResolvedT24} we can see that AGN likely do increase the calculated values of these two properties, especially for the barred distributions.

\begin{figure}
	\centering
    \includegraphics[width=0.45\textwidth,trim=0cm 0cm 0cm 1.0cm,clip]{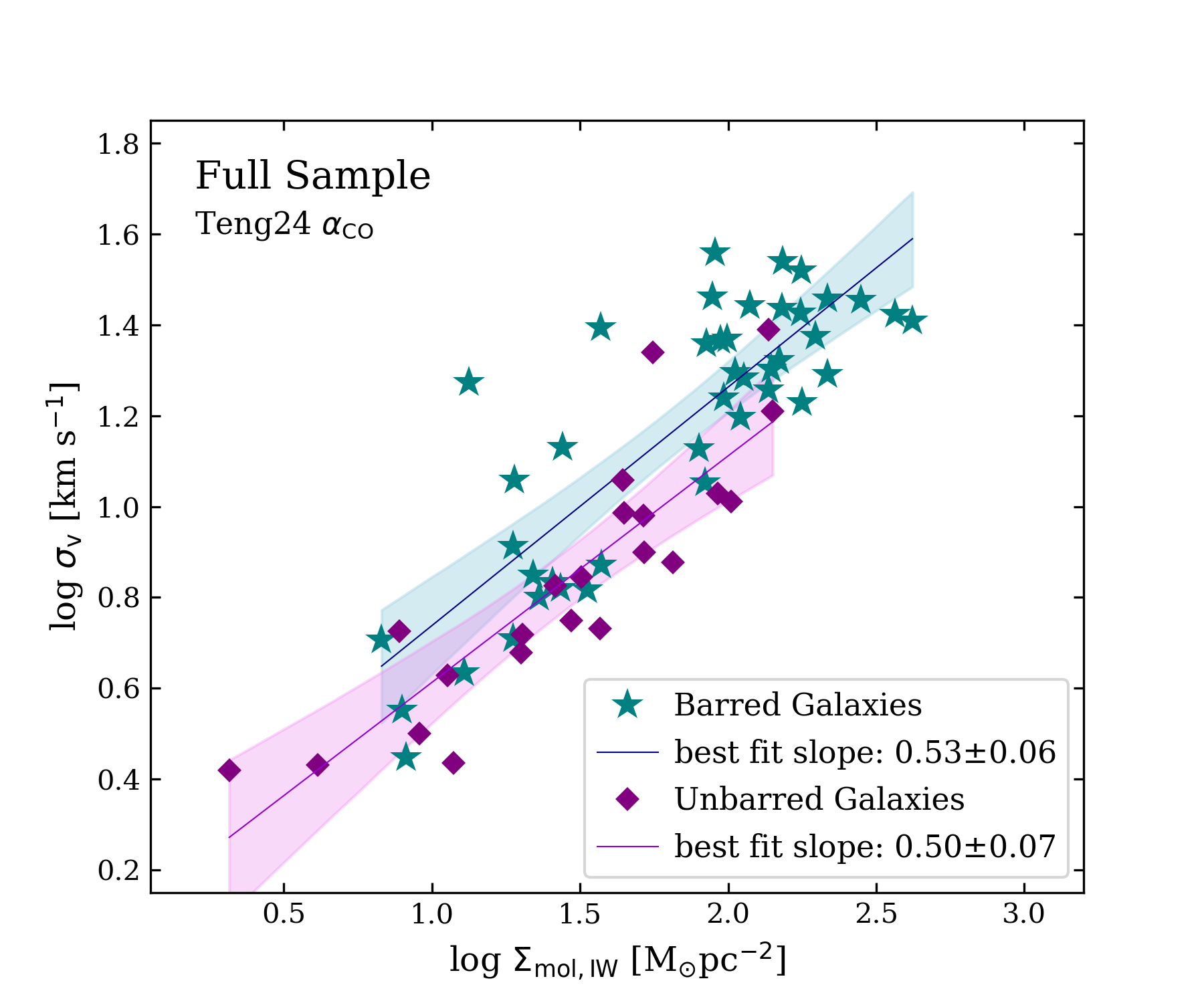} \\
    \includegraphics[width=0.45\textwidth,trim=0cm 0cm 0cm 1.0cm,clip]{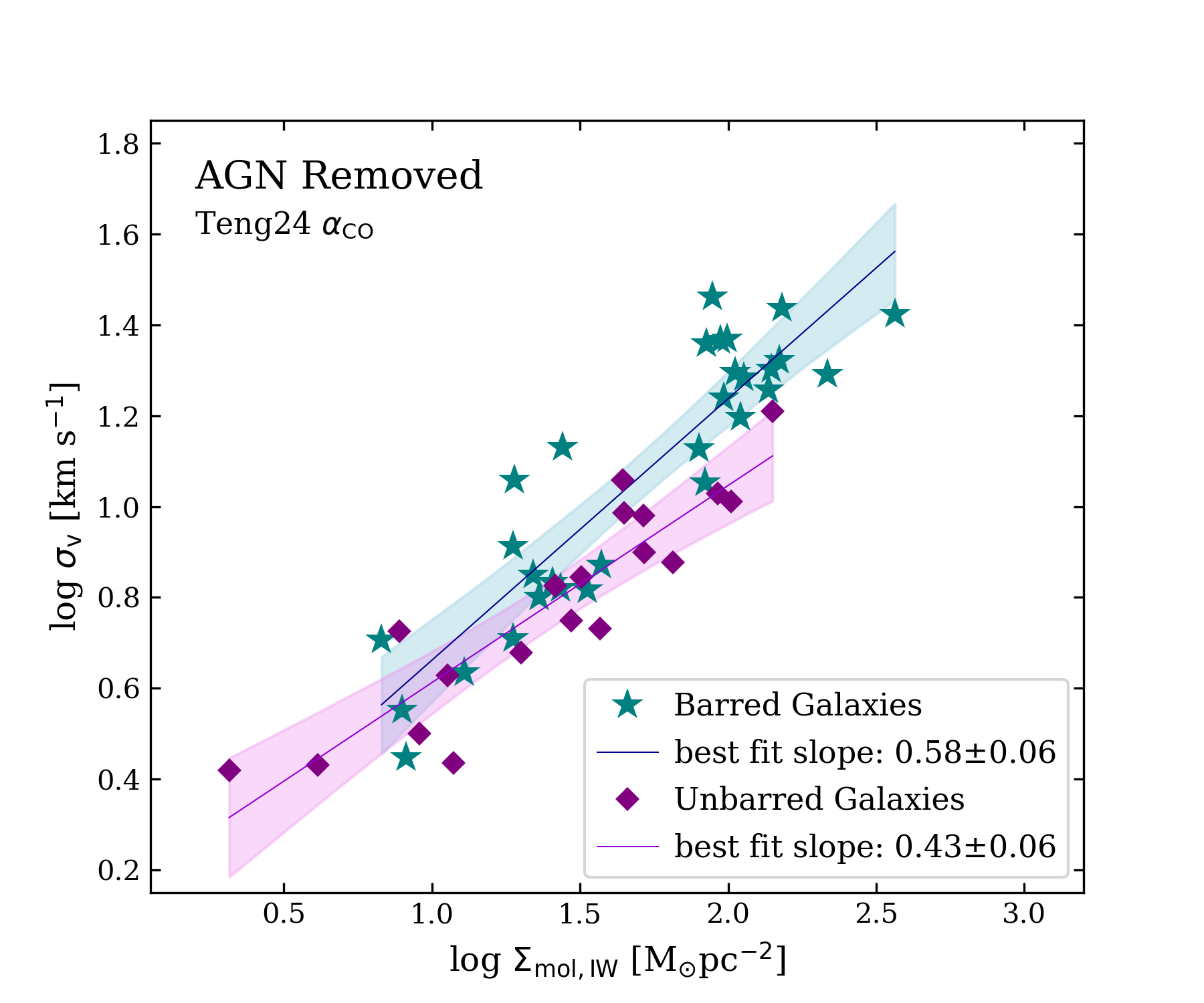}

    \caption{Barred versus unbarred resolved galaxy properties in the central region. [Top] $\sigma_{\rm{v}}$ as a function of $\Sigma_{\rm{mol,IW}}$ for the full sample, and [bottom] $\sigma_{\rm{v}}$ as a function of $\Sigma_{\rm{mol,IW}}$ for the sample with galaxies with an AGN removed. Best fit lines generated with Linmix with slope and intercept uncertainties shown in shaded regions. Even once galaxies with an AGN are removed, barred galaxies still show increased $\sigma_{\rm{v}}$ in the central region.}
    \label{fig:ResProp_Linmix_vdispSigmolIW_nAGN_T24}
\end{figure}

\begin{figure*}
	\centering
	\includegraphics[width=0.495\textwidth,trim=0cm 0cm 0.5cm 1.0cm,clip]{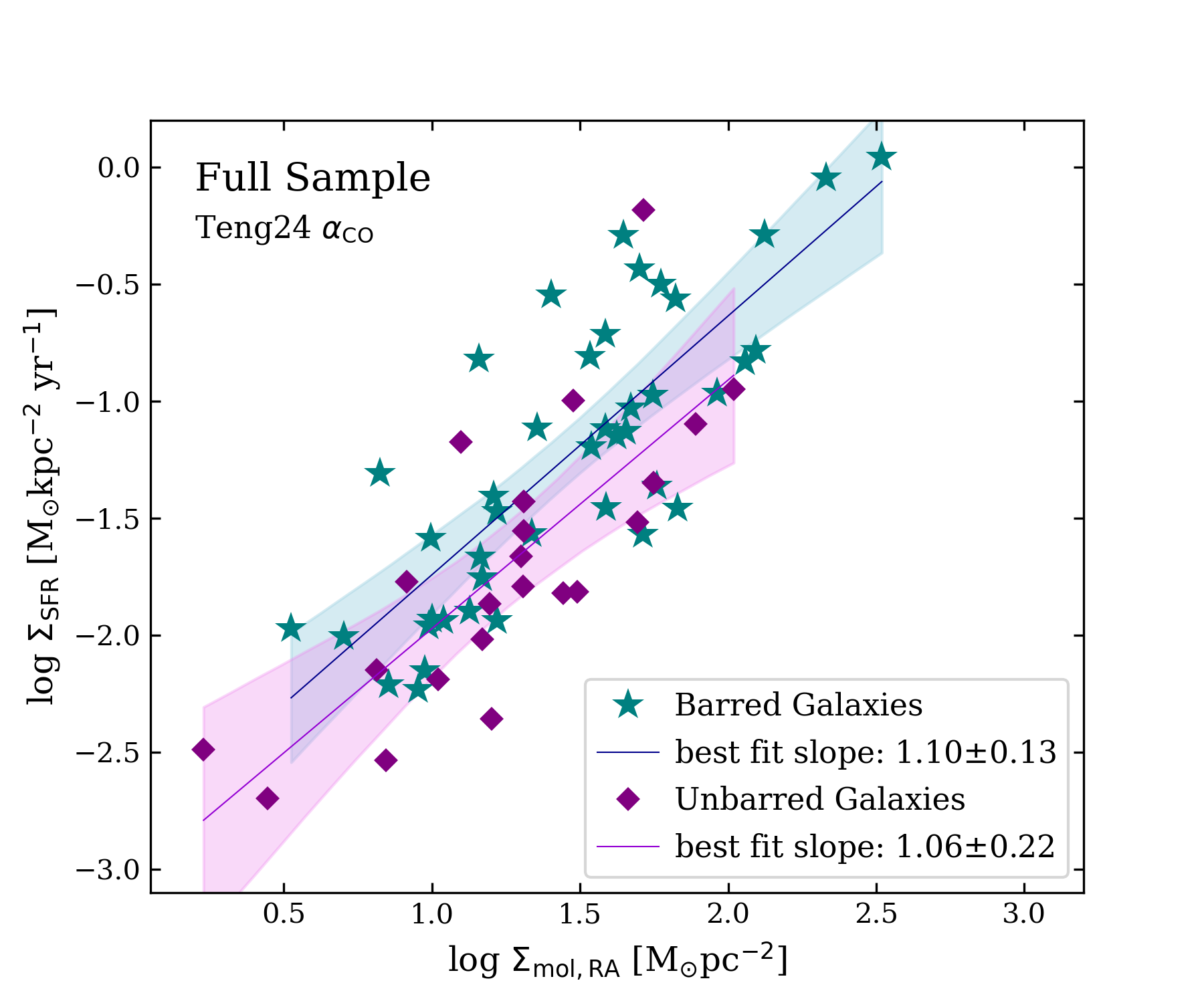}
    \includegraphics[width=0.495\textwidth,trim=0cm 0cm 0.5cm 1.0cm,clip]{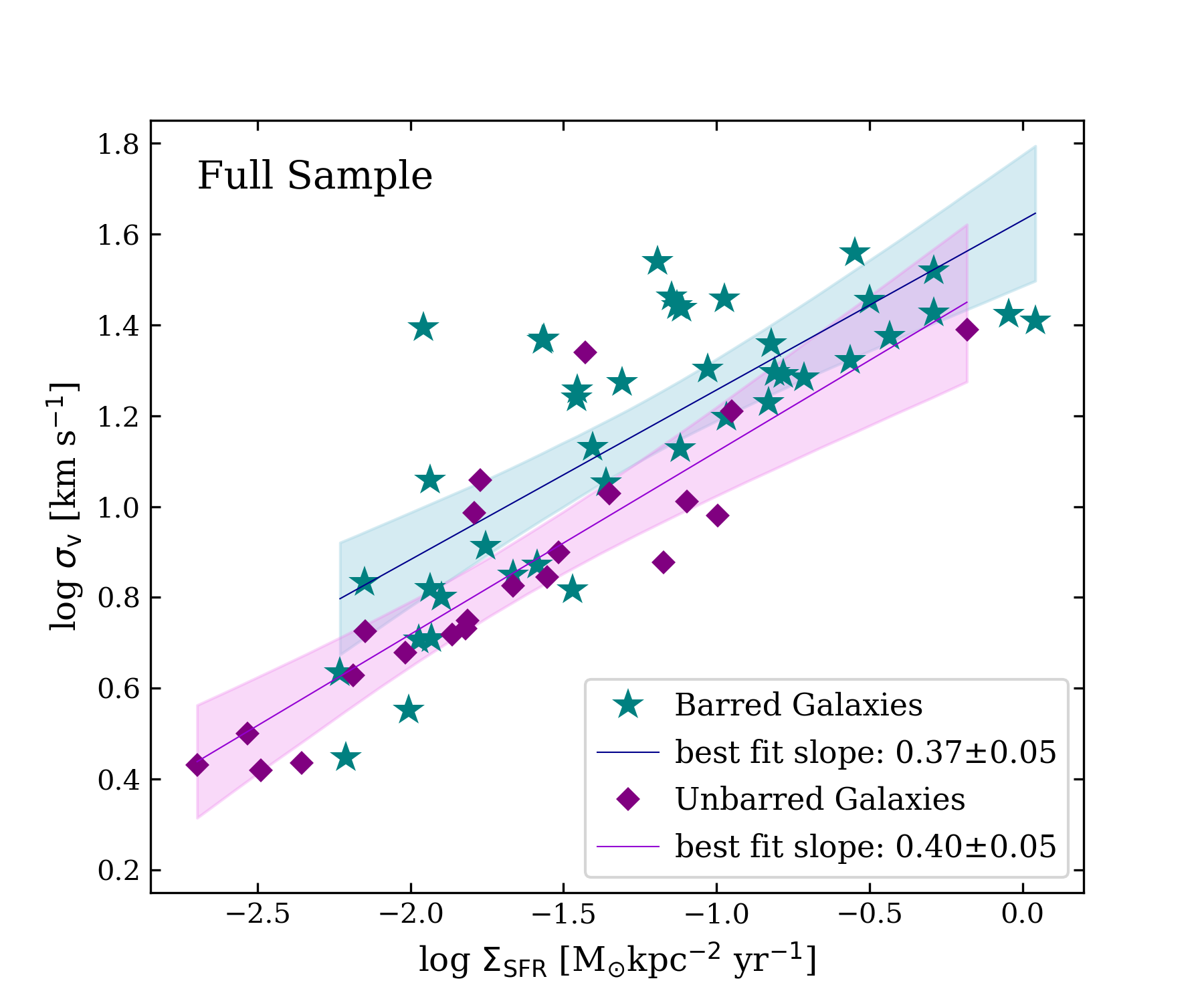}
    \caption{[Left] The Kennicutt-Schmidt relation $\Sigma_{\rm{SFR}}$ as a function of $\Sigma_{\rm{mol,RA}}$ for barred versus unbarred resolved molecular gas observations of the central region. The two samples agree very well. [Right] $\sigma_{\rm{v}}$ as a function of $\Sigma_{\rm{SFR}}$ for barred versus unbarred resolved central galaxy properties. Barred galaxies show somewhat increased $\sigma_{\rm{v}}$ in the central region. Best fit lines generated with Linmix with slope and intercept uncertainties shown in shaded regions. }
    \label{fig:ResProp_Linmix_KS-vdispSigSFR_T24}
\end{figure*}

\subsection{Correlations between physical quantities}  \label{subsec:2Linmix}

We consider the properties in a two-dimensional context by performing a linear regression analysis on three different combinations of physical quantities. We use Linmix (a hierarchical Bayesian python software package) to generate the most probable best fit line for the input data. The three comparisons we analyze are: $\sigma_{\rm{v}}$ vs $\Sigma_{\rm{mol,IW}}$; $\Sigma_{\rm{SFR}}$ vs $\Sigma_{\rm{mol,RA}}$; and $\sigma_{\rm{v}}$ vs $\Sigma_{\rm{SFR}}$.

Figures \ref{fig:ResProp_Linmix_vdispSigmolIW_nAGN_T24} and \ref{fig:ResProp_Linmix_KS-vdispSigSFR_T24} show the linear regression results, where the shaded regions indicate the slope and intercept uncertainties. These uncertainties are calculated as the 95\% confidence intervals for the slopes and intercepts. The measurement uncertainties for the hexagonal pixel in the centre of each galaxy are quite small so error bars do not show up for individual points on these plots. The linear regression results are listed in Table \ref{tab:LinmixResultsTableT24}.\footnote{We looked at the two different ways of averaging $\Sigma_{\rm{mol}}$ over the hexagonal pixel. The barred and unbarred galaxies follow similar relations in both cases.}

The comparison between the molecular gas surface density and velocity dispersion (Figure \ref{fig:ResProp_Linmix_vdispSigmolIW_nAGN_T24}) is important because it was in this distribution that \cite{SunJiayi2020} saw the higher values for barred galaxy centres.\footnote{In their analysis, \cite{SunJiayi2020} used 150 pc pixel data directly without any averaging into larger pixels.} Figure \ref{fig:ResProp_Linmix_vdispSigmolIW_nAGN_T24} and Table \ref{tab:ADbothSamplesT24} indicate that the significant difference between barred and unbarred galaxies for $\sigma_{\rm{v}}$ and $\Sigma_{\rm{mol,IW}}$ is partly due to the barred galaxy population having larger average values of both quantities (also seen in Figure \ref{fig:boxResolvedT24}). This comparison is consistent with the expectation that the inflow of gas along the bar causes extra turbulence as well as higher shear and/or non-circular motions.

Figure \ref{fig:ResProp_Linmix_vdispSigmolIW_nAGN_T24} (lower) shows the results of the Linmix linear regression for the comparison between $\Sigma_{\rm{mol,IW}}$ and $\sigma_{\rm{v}}$ where galaxies with an AGN have been removed. The slopes for both the barred and unbarred distributions change slightly, with barred galaxies having a steeper slope and unbarred galaxies having a shallower slope. The barred galaxies still have clearly higher velocity dispersions. The Linmix fit results for the three pairs of quantities are given in Table \ref{tab:LinmixResultsTableT24}.

Since $\Sigma_{\rm{SFR}}$ is region averaged over the hexagonal pixel we use the region averaged $\Sigma_{\rm{mol,RA}}$ to plot the Kennicutt-Schmidt (KS) relation for the full sample shown in Figure \ref{fig:ResProp_Linmix_KS-vdispSigSFR_T24} (left). We find a slope of $1.10\pm0.13$ for the barred sample and $1.06\pm0.22$ for the unbarred sample and very little offset between the two distributions. These slopes are similar to the slope of 0.91$\pm0.08$ found by \cite{Jimenez-Donaire2023} using data at 1.2 kpc resolution for the disks of Virgo Cluster galaxies. The slope of close to 1.0 also agrees with a study of resolved gas in field galaxies using the HERACLES sample \citep{Brown2023}. Our result suggests that the resolved KS relation is the same for centres as for the whole galaxy. Figure \ref{fig:ResProp_Linmix_KS-vdispSigSFR_T24} (left) also shows that $t_{\rm{dep}}$ is at the low end of the range for molecular gas in spiral galaxies of approximately $1-3$ Gyr  (\citealt{Querejeta2021}; see also Table \ref{tab:ResolvedPropT24}).

\begin{deluxetable*}{llcc}
\tablenum{4}
\tablecaption{Fits to pairs of physical quantities 
\label{tab:LinmixResultsTableT24}}
\tablewidth{0pt}
\tablehead{
\multicolumn{1}{l}{Sample} & \colhead{Property comparison} & \colhead{Barred} & \colhead{Unbarred}
} 
\startdata
Full Sample & $\sigma_{\rm{v}}$ vs $\Sigma_{\rm{mol,IW}}$
    & y = (0.53$\pm$0.06)x + (0.21$\pm$0.11) 
    & y = (0.50$\pm$0.07)x + (0.12$\pm$0.11) \\ 
& $\Sigma_{\rm{SFR}}$ vs $\Sigma_{\rm{mol,RA}}$
    & y = (1.10$\pm$0.13)x -- (2.85$\pm$0.20)
    & y = (1.06$\pm$0.22)x -- (3.03$\pm$0.29) \\
& $\sigma_{\rm{v}}$ vs $\Sigma_{\rm{SFR}}$
    & y = (0.37$\pm$0.05)x + (1.63$\pm$0.07) 
    & y = (0.40$\pm$0.05)x + (1.52$\pm$0.10) \\
\hline
AGN Removed & $\sigma_{\rm{v}}$ vs $\Sigma_{\rm{mol,IW}}$
    & y = (0.58$\pm$0.06)x + (0.09$\pm$0.10) 
    & y = (0.43$\pm$0.06)x + (0.18$\pm$0.08) \\ 
& $\Sigma_{\rm{SFR}}$ vs $\Sigma_{\rm{mol,RA}}$
    & y = (1.04$\pm$0.16)x -- (2.84$\pm$0.23)
    & y = (0.93$\pm$0.19)x -- (2.93$\pm$0.24) \\
& $\sigma_{\rm{v}}$ vs $\Sigma_{\rm{SFR}}$
    & y = (0.43$\pm$0.07)x + (1.68$\pm$0.10) 
    & y = (0.39$\pm$0.06)x + (1.49$\pm$0.10) \\
\enddata
\tablecomments{Statistical results of Linmix linear regression comparing 42 barred and 22 unbarred galaxies in the full sample, and 30 barred and 19 unbarred galaxies with no AGN for each comparison. In each case the resulting Linmix best fit slopes (with uncertainty) and intercepts (with uncertainty) are listed. Gas calculations are done using the linewidth-based $\alpha_{\rm{CO}}$ from \cite{Teng2024}.}
\end{deluxetable*}

Figure \ref{fig:ResProp_Linmix_KS-vdispSigSFR_T24} (right) shows $\sigma_{\rm{v}}$ as a function of $\Sigma_{\rm{SFR}}$ (again for the full sample) which is not as commonly plotted in the literature. The difference in offset between the two distributions implies that $\sigma_{\rm{v}}$ is higher for the barred galaxy centres. Higher velocity dispersion can indicate more turbulence and/or non-circular motions in the gas which can be driven by and ultimately impact star formation.

\subsection{Discussion and Implications}
\label{sec:implications}

The implications of the observed higher linewidths in the barred galaxy centres requires a thoughtful approach, as there are many possible contributions to the velocity dispersion in these extreme central environments.
One contribution to linewidth when observing molecular clouds is the cloud's own internal turbulence \citep[e.g.,][]{2017MivilleDeschenes}. If there is more than one cloud within the pixel, additional broadening can come from the cloud-cloud velocity dispersion, which is the dispersion in the mean velocities of clouds in the population \citep[e.g.,][]{Wilson2011JCMT}. There can be shear across the cloud
due to the gradient of the mean rotation curve of the galaxy \citep[e.g.][]{Choi2023}. Other bulk motions of the gas such as the flow along the bar \citep[e.g.][]{Hogarth2024} could also contribute shear that can broaden the lines.

Any velocity gradients across the beam due to the rotation curve of the galaxy or large-scale flows can also produce broadening via the non-physical mechanism of beam smearing, e.g. line broadening that would not be present if the resolution of the data was sufficiently high. Although the 150 pc resolution of the PHANGS-ALMA observations reaches close to the physical sizes of the largest molecular clouds, higher resolution observations in some galaxy centres have shown that significant velocity gradients can exist on this scale \citep[e.g.][]{North2019,Liu2021}. However, as discussed in Section~\ref{subsec:data}, the similar mass distributions between the barred and unbarred galaxy samples means that we do not expect beam smearing to affect the two samples differently.

We find that for the full sample, as well as the sample with AGN removed, the velocity dispersion is statistically different between the barred and unbarred galaxies. One possibility is that either the internal turbulence of individual clouds or the cloud-cloud velocity dispersion of the population as a whole could be higher in the barred galaxy centres. Another possible factor could be an increased shear experienced by clouds in barred galaxy centers. \citet{Rosolowsky2021MNRAS} analyzed individual clouds from PHANGS data identified using pycprops (instead of the pixel-based analysis) and find that clouds in galaxies with bars have larger linewidths towards the central region. They attribute this increase to non-circular motions of the clouds, which could be streaming motions along the bar, stretching or shearing of individual clouds, or clouds in the same orbit overlapping along the line of sight.

In addition, the consistent slope for both barred and unbarred galaxies seen in Figure \ref{fig:ResProp_Linmix_KS-vdispSigSFR_T24} (right) suggests that any extra $\sigma_{\rm{v}}$ due to the bar is not constant but has to increase with $\Sigma_{\rm{SFR}}$. However, whether the increased linewidths in barred galaxy centres are due to increased internal cloud velocity dispersion (internal turbulence), increased cloud-cloud velocity dispersion (population turbulence), or shear affecting individual clouds cannot be determined with the data in hand. 

Along with higher linewidths, we find statistically lower $t_{\rm{dep}}$ in barred galaxy centres whether or not there are AGN in the sample. This raises the question of whether the difference in $t_{\rm{dep}}$ can help us constrain the physical origin of the different linewidths. The depletion time is calculated as the ratio of two ``lower resolution" measurements (the region averaged $\Sigma_{\rm{mol,RA}}$ and $\Sigma_{\rm{SFR}}$). Using the linewidth-dependent $\alpha_{\rm{CO}}$ prescription, $\Sigma_{\rm{mol,RA}}$ and therefore $t_{\rm{dep}}$ depend roughly inversely on $\sigma_{\rm{v}}$, 
\begin{equation}
    t_{\rm{dep}} = \frac{\Sigma_{\rm{mol,RA}}}{\Sigma_{\rm{SFR}}} = \frac{I_{\rm{CO}} \cos {\it i} \ 10^{1.05}}{\Sigma_{\rm{SFR}} \ R_{21} \ \sigma_{\rm{v}}^{0.81}}.
\end{equation}

\noindent For the full sample, we find that $\Sigma_{\rm{mol,RA}}$ is not different for the barred or unbarred populations, while $\Sigma_{\rm{SFR}}$ and $\sigma_{\rm{v}}$ are higher in barred galaxies, as is $I_{\rm{CO}} \cos i$. The fact that $\Sigma_{\rm{SFR}}$, $I_{\rm{CO}} \cos i$, and $\sigma_{\rm{v}}$ are all higher in barred galaxies naturally explains the lower depletion times.
 
However, when AGN are removed from the sample, only $t_{\rm{dep}}$ and $\sigma_{\rm{v}}$ show differences between the barred and unbarred populations. Equation (4) then suggests that the inverse dependence of $\alpha_{\rm{CO}}$ on $\sigma_{\rm{v}}$ is driving the lower $t_{\rm{dep}}$ in barred centres. We do not see the lower $t_{\rm{dep}}$ for bars when using the metallicity-dependent $\alpha_{\rm{CO}}$ (see Appendix \ref{AppendixB}), which shows that the differences in $t_{\rm{dep}}$ for barred and unbarred galaxies depend on the choice of $\alpha_{\rm{CO}}$.

The $t_{\rm{dep}}$ for barred galaxy centres is shorter (as low as 100 Myr) than the range of approximately $1-3$ Gyr for the molecular gas in unbarred centres and spiral arms \citep{Querejeta2021}. The lower $t_{\rm{dep}}$ is in agreement with the results of \citet{Teng2024}, who compared 4 different $\alpha_{\rm{CO}}$ prescriptions and found that the linewidth-dependent $\alpha_{\rm{CO}}$ resulted in the lowest $t_{\rm{dep}}$ in barred galaxy centres. We note that a recent study \citep{Hogarth2024} found that $t_{\rm{dep}}$ is higher (i.e. SFE is suppressed) in bars which are actively driving a radial flow of gas. This result aligns with studies comparing different parts of the bar \citep{Maeda2023SFgasrichBars}, which found that gas along the length of the bar has higher $t_{\rm{dep}}$ (lower SFE), likely due to large dynamical effects such as shocks and shear. In contrast, the ends and central regions of the bar are locations where $t_{\rm{dep}}$ is comparable to or smaller than values in the disk (implying similar or higher SFE). 

Given the variations in $t_{\rm{dep}}$ as a function of position along the bar seen in \cite{Maeda2023SFgasrichBars}, we examined whether $t_{\rm{dep}}$  depends on  the fraction of the bar included in the central 1.5 kpc region. In our sample of 42 barred galaxies, only 2 have bars that are completely contained inside the 1.5 kpc aperture. For the majority of the galaxies in the sample (70$\%$) the aperture encompasses less than 30$\%$ of the bar. At the 1.5 kpc resolution of our data, we find no dependence of $t_{\rm{dep}}$ on the fraction of the bar inside the aperture.


\section{Conclusions} \label{sec:conclusions}

In this paper we have carried out a statistical analysis of the central molecular gas and star formation properties for barred and unbarred galaxies using high resolution molecular gas data and medium resolution SFR data from PHANGS. We have compared five key properties in the central regions: $\Sigma_{\rm{mol,RA}}$, $\Sigma_{\rm{mol,IW}}$, $\sigma_{\rm{v}}$, $\Sigma_{\rm{SFR}}$, and $t_{\rm{dep}}$. $\Sigma_{\rm{mol,RA}}$ and $\Sigma_{\rm{mol,IW}}$ represent two different ways to average molecular gas surface density over the hexagonal pixel. 

\begin{enumerate}
    \item Anderson-Darling tests show statistically significant differences between barred and unbarred galaxy centres for all quantities in the full sample except the region averaged $\Sigma_{\rm{mol,RA}}$. The same statistical tests for a subsample of the galaxies without an AGN show that  $\sigma_{\rm{v}}$ and $t_{\rm{dep}}$ remain clearly statistically different for the barred and unbarred galaxy centres.
    \item We performed a linear regression analysis (using Linmix) on three different pairs of properties: $\sigma_{\rm{v}}$ vs $\Sigma_{\rm{mol,IW}}$; $\Sigma_{\rm{SFR}}$ vs $\Sigma_{\rm{mol,RA}}$; and $\sigma_{\rm{v}}$ vs $\Sigma_{\rm{SFR}}$. The higher $\sigma_{\rm{v}}$ in the barred galaxy sample seen in the plot of $\sigma_{\rm{v}}$ vs $\Sigma_{\rm{mol,IW}}$ suggests the presence of extra turbulence, higher shear and/or non-circular motions, likely due to inflow of gas along the bar. Even when AGN are removed from the sample, $\sigma_{\rm{v}}$ continues to be higher in barred centres. The exact physical mechanism at work (internal turbulence, population turbulence, or shear) would require further analysis of higher resolution data. The offset between the barred and unbarred samples in the $\sigma_{\rm{v}}$ vs $\Sigma_{\rm{SFR}}$ relation with consistent slope suggests that star formation must increase with the increased linewidths in the gas found in barred galaxy centres.
    \item We find that barred centres have shorter $t_{\rm{dep}}$, with the shortest being on the order of 100 Myr, while unbarred centres have $t_{\rm{dep}}$ consistent with the expected range of approximately $1-3$ Gyr seen for the molecular gas in spiral galaxies \citep{Querejeta2021}. We find no trend of $t_{\rm{dep}}$ with the fraction of the bar contained in the central 1.5 kpc aperture. We note that $t_{\rm{dep}}$ is very sensitive to the choice of $\alpha_{\rm{CO}}$ prescription and the difference between barred and unbarred galaxies is not seen with a metallicity-dependent $\alpha_{\rm{CO}}$ \citep[see also][]{Teng2024}. 
    \end{enumerate}

Two of the properties that show the strongest evidence for differences between the two samples ($\Sigma_{\rm{mol,IW}}$ and $\sigma_{\rm{v}}$) are the two quantities that are intensity-weighted over the hexagonal pixel and so require the availability of high resolution measurements. This shows that the availability of higher resolution data is important for the statistics of the gas properties. The SFR data used in this paper were medium resolution; future work will make use of data from a new large program, Multiphase Astrophysics to Unveil the Virgo Environment (MAUVE), which will provide $\sim$100 pc scale ionized gas excitation and SFR maps.


\begin{acknowledgments}
We thank the referee for thoughtful comments that resulted in significant changes to our approach in this paper. We are grateful to Dr. Jiayi Sun and other members of the PHANGS collaboration whose work motivated this project and for their extensive support during the analysis. ALMA is a partnership of ESO (representing its member states), NSF (USA), and NINS (Japan), together with NRC (Canada), MOST and ASIAA (Taiwan), and KASI (Republic of Korea), in cooperation with the Republic of Chile. The Joint ALMA Observatory is operated by ESO, AUI/NRAO, and NAOJ. The National Radio Astronomy Observatory is a facility of the National Science Foundation operated under cooperative agreement by Associated Universities, Inc. 

This research has made use of the NASA/IPAC Extragalactic Database (NED) which is operated by the Jet Propulsion Laboratory, California Institute of Technology, under contract with the National Aeronautics and Space Administration. CDW acknowledges financial support from the Canada Council for the Arts through a Killam Research Fellowship. The research of CDW is supported by grants from the Natural Sciences and Engineering Research Council of Canada (NSERC) and the Canada Research Chairs program (CRC-2022-00184).

\end{acknowledgments}



\vspace{5mm}
\facilities{ALMA, GALEX, Spitzer, WISE}


\software{This research has made use of the following software packages: \texttt{ASTROPY}\footnote{https://www.astropy.org/}, a community-developed core \texttt{PYTHON} package for astronomy \citep{astropy2013, astropy2018, astropy2022}, \textsc{CASA}\footnote{https://casa.nrao.edu/} \citep{McMullin2007}, \texttt{MATPLOTLIB}\footnote{https://matplotlib.org/} \citep{Hunter2007}, \texttt{NUMPY}\footnote{https://numpy.org/} \citep{harris2020}, RStudio\footnote{https://posit.co/download/rstudio-desktop/} \citep{R2023}, \texttt{SciPy}\footnote{https://scipy.org/} \citep{scipy2020}, and \texttt{SPECTRAL-CUBE}\footnote{https://spectral-cube.readthedocs.io/en/latest/} \citep{Ginsburg2019}.}



\appendix


\section{Galaxy Properties}
\label{AppendixA}

In Table \ref{tab_galaxylist} we list the galaxies in our sample with classifications for whether or not the galaxy has a galactic bar or an AGN,  as well as the environment in which the galaxy is located (whether in a cluster, group or is isolated). If the galaxy has a bar we have listed the size of the semi-major axis in kpc, and the fraction of the bar that is contained within the central 1.5 kpc hexagonal pixel.

\startlongtable
\begin{deluxetable*}{lcccccl}
\tablenum{A1}
\tablecaption{PHANGS Galaxies used in this analysis \label{tab_galaxylist}}
\tablewidth{0pt}
\tablehead{
\colhead{Galaxy} & \colhead{Bar} & \colhead{Semi-major axis} & \colhead{Centre} & \colhead{AGN} & \colhead{Cluster} & \colhead{description} \\
\colhead{} & \colhead{[Y/N]} & \colhead{[kpc]} & \colhead{[ratio]} &  \colhead{[Y/N]} & \colhead{[Y/N]} & \colhead{-}
}
\decimalcolnumbers
\startdata
ESO097-013 & N & -      &  -      &   Y  & N  &  Isolated \\
IC1954     & Y & 0.48   &  1.00   &   N  & N  &  Group  \\
IC5273     & Y & 1.41   &  0.532  &   N  & N  &  Group (Sculptor)  \\
NGC0253\tablenotemark{\rm{*}} & Y & -    &  -   &  Y  & N  &  Group (Sculptor)  \\
NGC0300    & N & -      &  -      &   N  & N  &  Group (Sculptor)  \\
NGC0628    & N & -      &  -      &   N  & N  &  Isolated  \\
NGC0685    & Y & 1.54   &  0.486  &   N  & N  &  Pair Member  \\
NGC1087    & Y & 1.17   &  0.639  &   N  & N  &  Group  \\
NGC1097    & Y & 6.47   &  0.116  &   Y  & N  &  Group  \\
NGC1300    & Y & 10.2   &  0.0734 &   N  & Y  &  Cluster (Eridanus)   \\
NGC1317    & Y & 3.87   &  0.194  &   N  & Y  &  Cluster (Fornax)  \\
NGC1365    & Y & 7.93   &  0.0945 &   Y  & Y  &  Cluster (Fornax)  \\
NGC1385    & N & -      &  -      &   N  & Y  &  Cluster (Eridanus)  \\
NGC1433    & Y & 5.37   &  0.140  &   N  & N  &  Group  \\
NGC1511    & N & -      &  -      &   N  & N  &  Group  \\
NGC1512    & Y & 5.82   &  0.129  &   N  & N  &  Group  \\
NGC1546    & N & -      &  -      &   N  & N  &  Group (Dorado)  \\
NGC1559    & Y & 1.07   &  0.704  &   N  & N  &  Group (Dorado)  \\
NGC1566    & Y & 3.12   &  0.240  &   Y  & N  &  Group (Dorado)  \\
NGC1637    & Y & 0.95   &  0.788  &   N  & N  &  -  \\
NGC1792    & N & -      &  -      &   N  & N  &  Group  \\
NGC2090    & N & -      &  -      &   N  & N  &  -  \\
NGC2283    & Y & 0.605  &  1.00   &   N  & N  &  Pair Member  \\
NGC2566    & Y & 6.89   &  0.109  &   N  & N  &  Group  \\
NGC2835    & Y & 1.03   &  0.729  &   N  & N  &  Group  \\
NGC2903    & Y & 2.80   &  0.268  &   N  & N  &  Isolated  \\
NGC2997    & N & -      &  -      &   N  & N  &  Group  \\
NGC3137    & N & -      &  -      &   N  & N  &  Group  \\
NGC3351    & Y & 2.49   &  0.301  &   N  & N  &  Group (Leo)  \\
NGC3507    & Y & 2.58   &  0.290  &   N  & N  &  Group  \\
NGC3511    & Y & 0.907  &  0.827  &   N  & N  &  Pair Member  \\
NGC3521    & N & -      &  -      &   N  & N  &  Isolated  \\
NGC3596    & N & -      &  -      &   N  & N  &  Isolated  \\
NGC3621    & N & -      &  -      &   Y  & N  &  Isolated  \\
NGC3626    & Y & 1.92   &  0.391  &   N  & N  &  Group  \\
NGC3627    & Y & 3.40   &  0.220  &   Y  & N  &  Group  \\
NGC4254    & N & -      &  -      &   N  & Y  &  Cluster (Virgo)  \\
NGC4293    & Y & 6.06   &  0.124  &   N  & Y  &  Cluster (Virgo)  \\
NGC4298    & N & -      &  -      &   N  & Y  &  Cluster (Virgo)  \\
NGC4303    & Y & 3.08   &  0.243  &   Y  & Y  &  Cluster (Virgo)  \\
NGC4321    & Y & 4.36   &  0.172  &   N  & Y  &  Cluster (Virgo)  \\
NGC4457    & Y & 2.47   &  0.304  &   Y  & Y  &  Cluster (Virgo)  \\
NGC4496A   & Y & 1.69   &  0.444  &   -  & Y  &  Cluster (Virgo)  \\
NGC4535    & Y & 2.88   &  0.260  &   N  & Y  &  Cluster (Virgo)  \\
NGC4536    & Y & 2.47   &  0.304  &   N  & Y  &  Cluster (Virgo)  \\
NGC4540    & Y & 1.47   &  0.509  &   N  & Y  &  Cluster (Virgo)  \\
NGC4548    & Y & 4.68   &  0.160  &   Y  & Y  &  Cluster (Virgo)  \\
NGC4569    & Y & 7.99   &  0.094  &   Y  &  Y  &  Cluster (Virgo)  \\
NGC4571    & N & -      &  -      &   N  & Y  &  Cluster (Virgo)  \\
NGC4689    & N & -      &  -      &   N  & Y  &  Cluster (Virgo)  \\
NGC4731    & Y & 3.46   &  0.217  &   N  & N  &  Group (Virgo Y)  \\
NGC4781    & Y & 1.35   &  0.556  &   N  & N  &  Group  \\
NGC4826    & N & -      &  -      &   Y  & N  &  Isolated  \\
NGC4941    & Y & 6.31   &  0.119  &   Y  & N  &  Group  \\
NGC4951    & N & -      &  -      &   N  & N  &  Group  \\
NGC5042    & N & -      &  -      &   N  & N  &  -  \\
NGC5068    & Y & 0.873  &  0.860  &   N  & N  &  Group  \\
NGC5134    & Y & 4.07   &  0.184  &   N  & N  &  Group  \\
NGC5248    & Y & 5.85   &  0.128  &   N  & N  &  Isolated  \\
NGC5530    & N & -      &  -      &   N  & N  &  -  \\
NGC5643    & Y & 3.09   &  0.243  &   Y  & N  &  -  \\
NGC6300    & Y & 2.29   &  0.328  &   Y  & N  &  Group  \\
NGC7456    & N & -      &  -      &   N  & N  &   Group  \\
NGC7496    & Y & 3.41   &  0.220  &   Y  & N  &   Group  \\
\hline 
\enddata
\tablecomments{(1) Galaxy name. (2) Shows if the galaxy has a bar. (3) The semi-major axis length of the bar if present. (4) The percentage of the bar that is contained within the central 1.5 kpc hexagonal pixel. Most bars extend beyond the central pixel. (5) Shows if the galaxy has an AGN. (6) Shows if the galaxy is in a cluster. (7) Description of where the galaxy is located, whether it is isolated, in a group, or in a cluster.}
\tablenotetext{*}{Galaxy is not part of the sample from \cite{Querejeta2021}, from which we obtained the bar size data.}
\end{deluxetable*}


\section{Analysis with metallicity dependent $\alpha_{\rm{CO}}$ }
\label{AppendixB}

In this appendix, we show the results of our analysis using the metallicity-dependent conversion factor ($\alpha_{\rm{CO}}$) from \cite{SunJiayi2020}, which is defined as
\begin{equation}
    \alpha_{\rm{CO}} = 4.35 \ Z'^{-1.6} [M_{\odot} \ \rm{pc}^{-2} (\rm{K} \ \rm{km \ s^{-1}})^{-1}],
\end{equation}

\noindent where $Z'$ is the local metallicity divided by $Z{_\odot}$. The metallicity was calculated by \citet{SunJiayi2022multi} analytically by inferring the metallicity at $R_e=1.68 r_{disk}$ (where $r_{disk}$ is the scale length of the stellar disk) from the global mass metallicity relationship after first applying an increase of 0.1 dex to the global mass. The metallicity for pixels at other radii was then calculated assuming a metallicity gradient of -0.1 dex/kpc.

We analyse the same five key properties as in the main text, the region-averaged $\Sigma_{\rm{mol,RA}}$, intensity-weighted mean $\Sigma_{\rm{mol,IW}}$, intensity-weighted mean $\sigma_{\rm{v}}$, region-averaged $\Sigma_{\rm{SFR}}$, and molecular gas depletion time ($t_{\rm{dep}}$) for the central hexagonal pixel. The samples of barred and unbarred galaxies are the same with 42 and 22 galaxies respectively for the full sample, and 30 barred and 19 unbarred galaxies in the subset with no AGN.

\begin{figure*}
	\centering
    \includegraphics[height=0.17\textwidth,trim=0cm 0cm 0cm 0cm,clip]{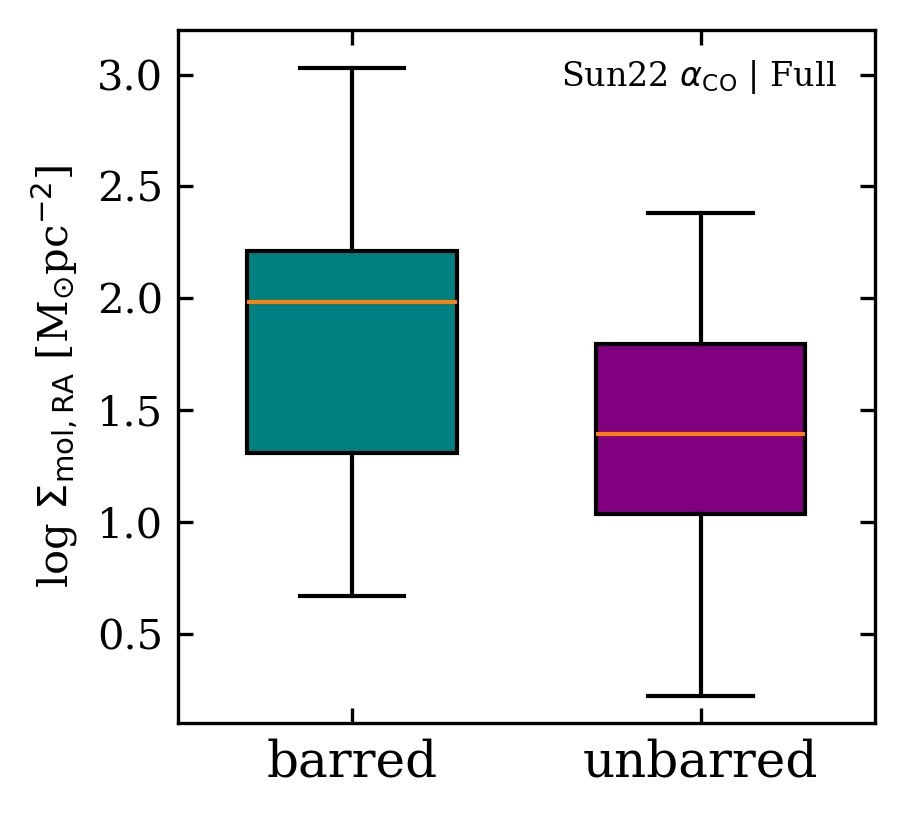} 
    \includegraphics[height=0.17\textwidth,trim=0cm 0cm 0cm 0cm,clip]{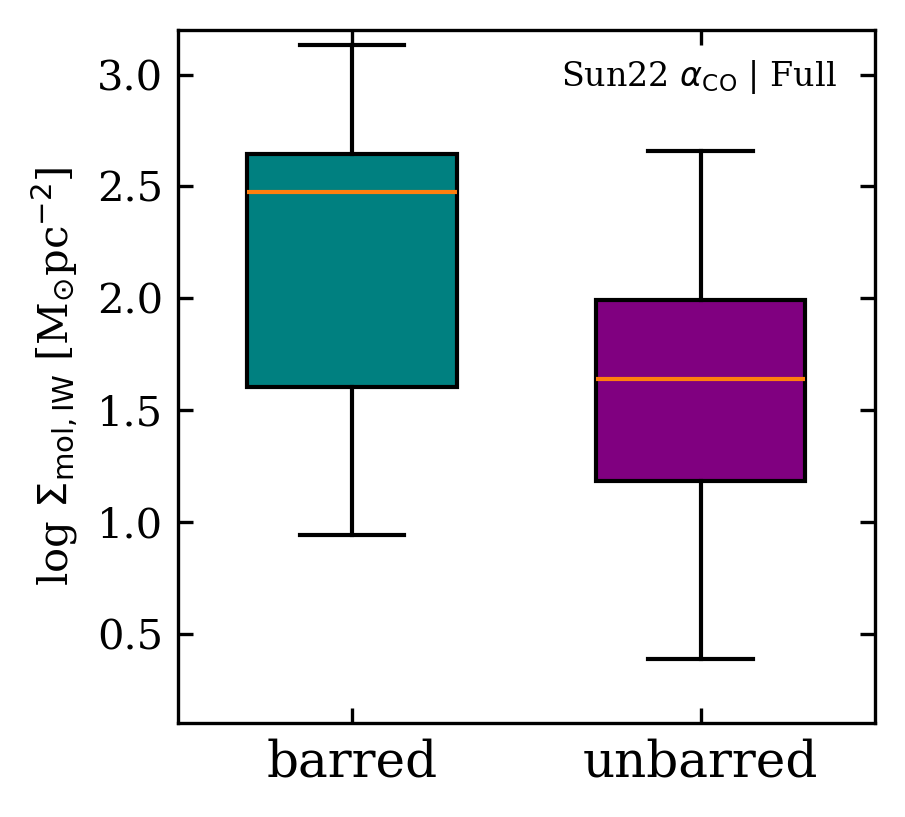} 
    \includegraphics[height=0.17\textwidth,trim=0cm 0cm 0cm 0cm,clip]{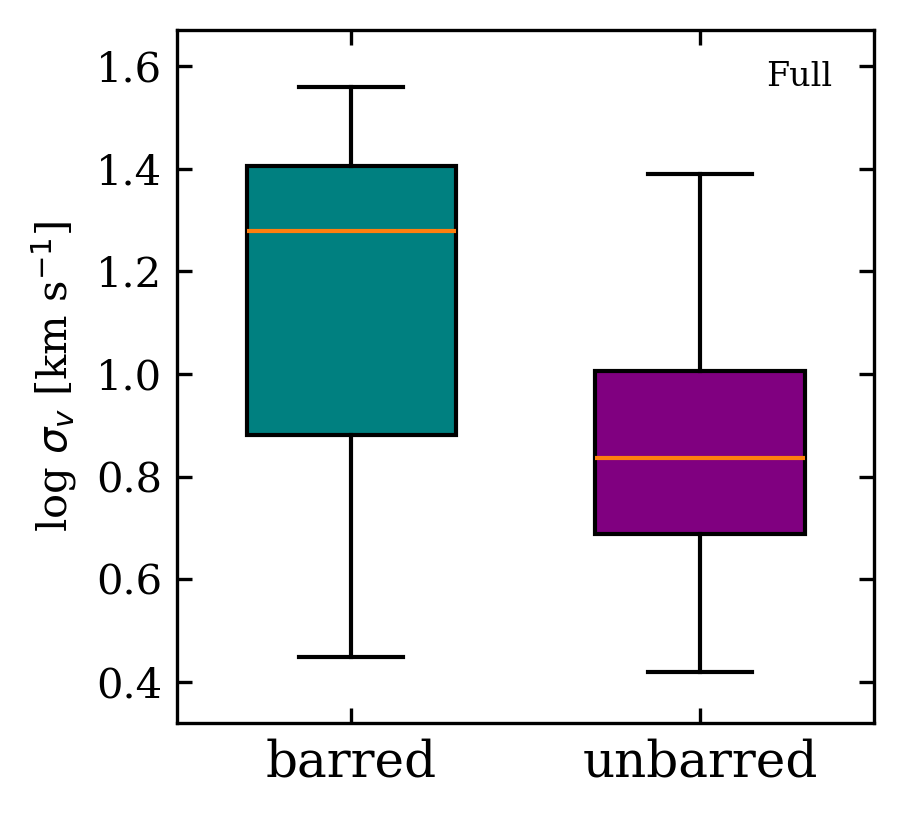}
    \includegraphics[height=0.17\textwidth,trim=0cm 0cm 0cm 0cm,clip]{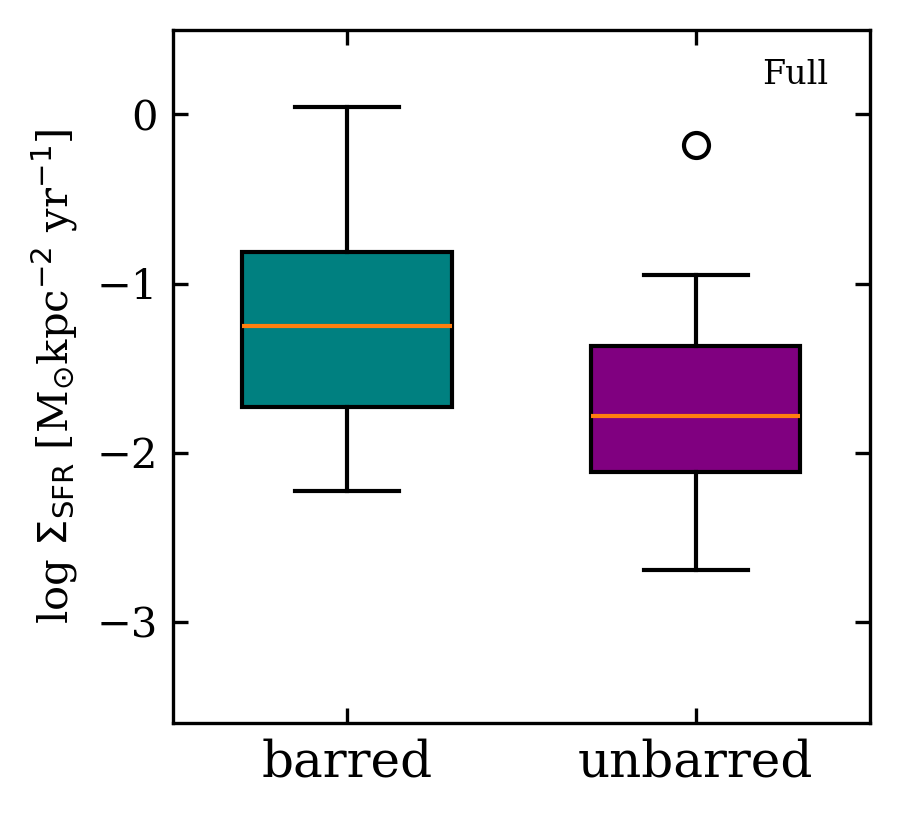}
    \includegraphics[height=0.17\textwidth,trim=0cm 0cm 0cm 0cm,clip]{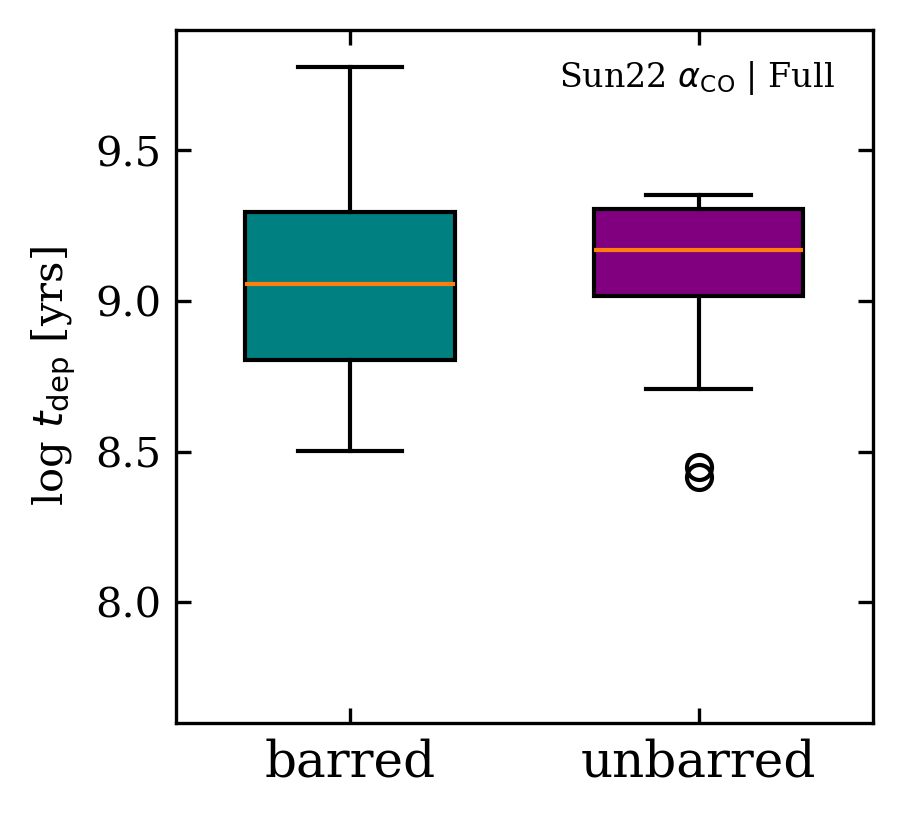} \\
    \includegraphics[height=0.17\textwidth,trim=0cm 0cm 0cm 0cm,clip]{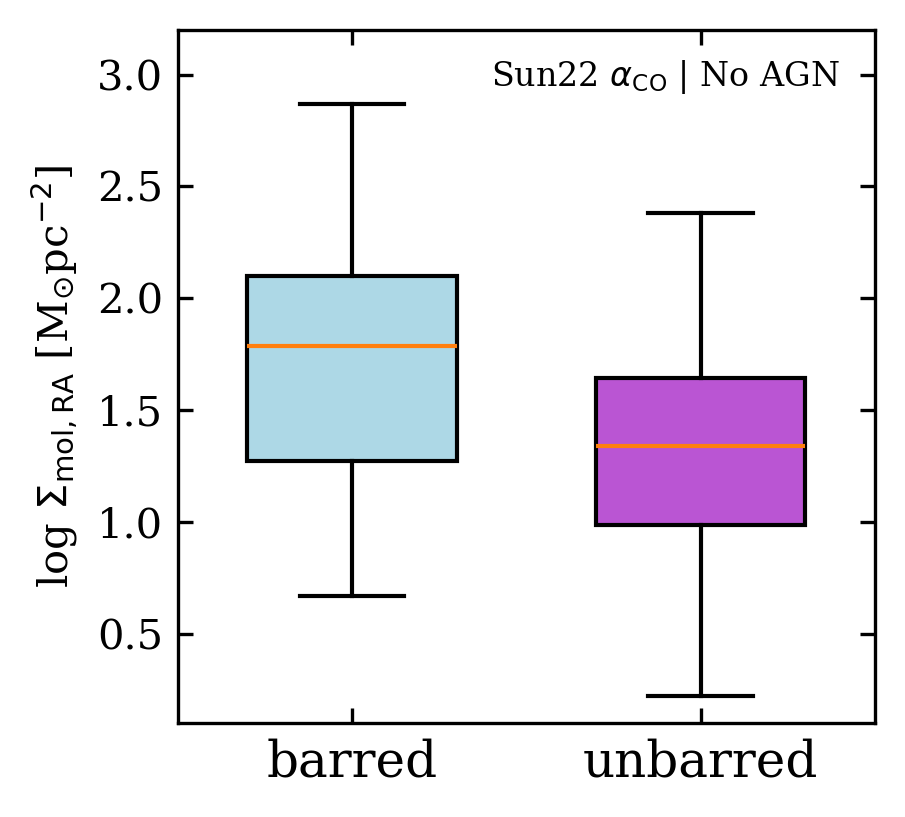} 
    \includegraphics[height=0.17\textwidth,trim=0cm 0cm 0cm 0cm,clip]{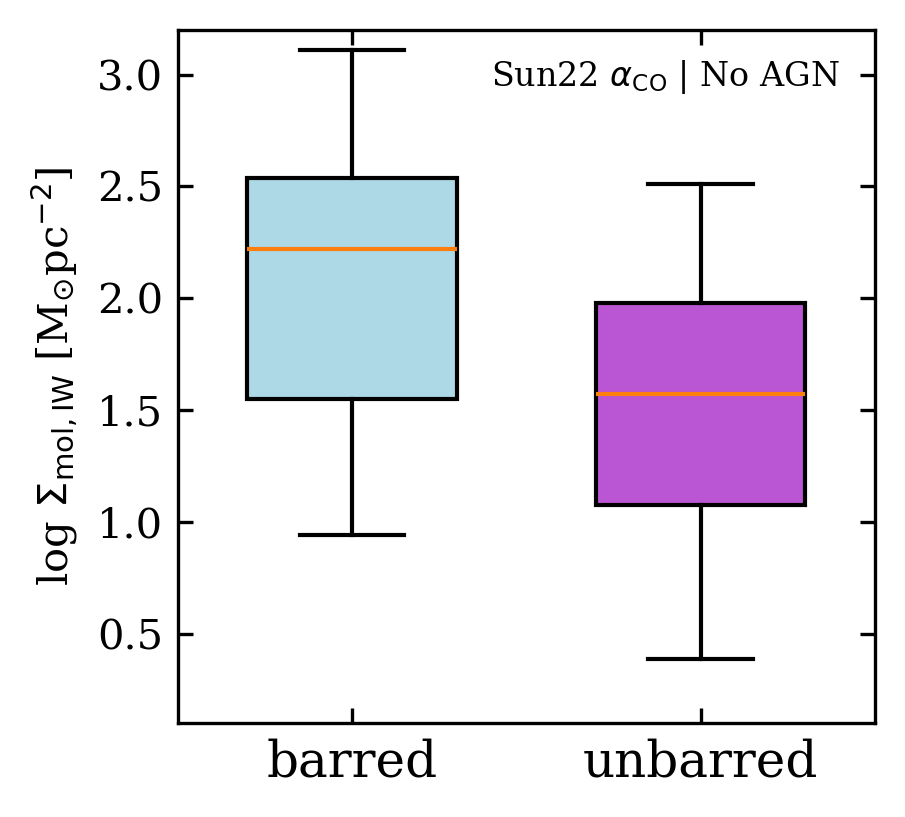} 
    \includegraphics[height=0.17\textwidth,trim=0cm 0cm 0cm 0cm,clip]{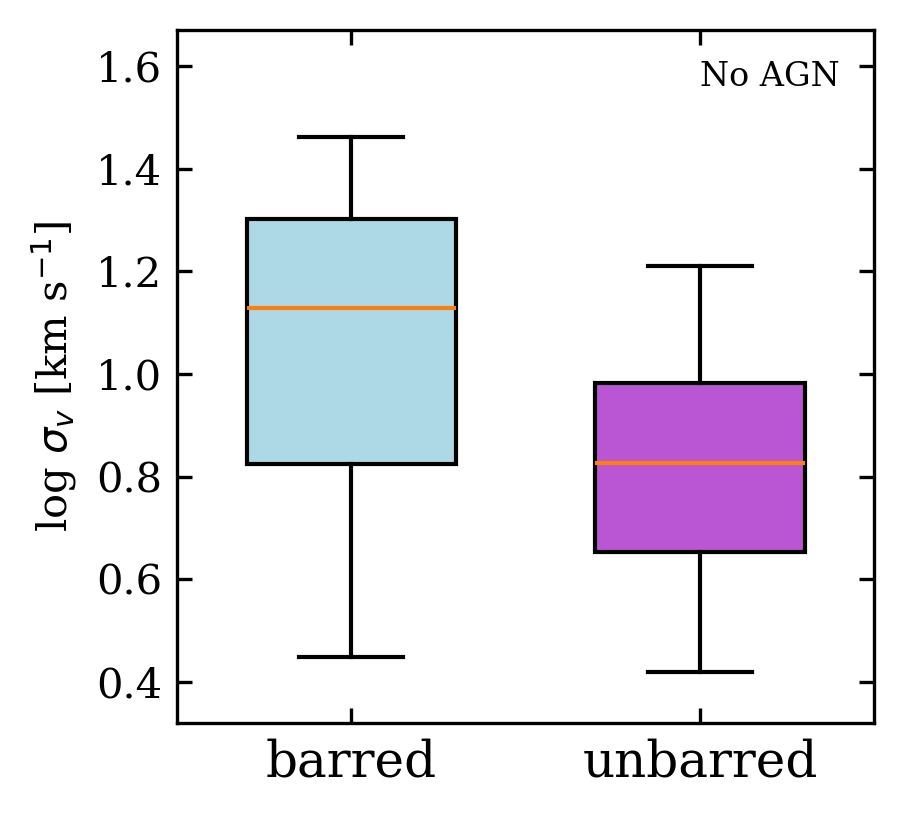}
    \includegraphics[height=0.17\textwidth,trim=0cm 0cm 0cm 0cm,clip]{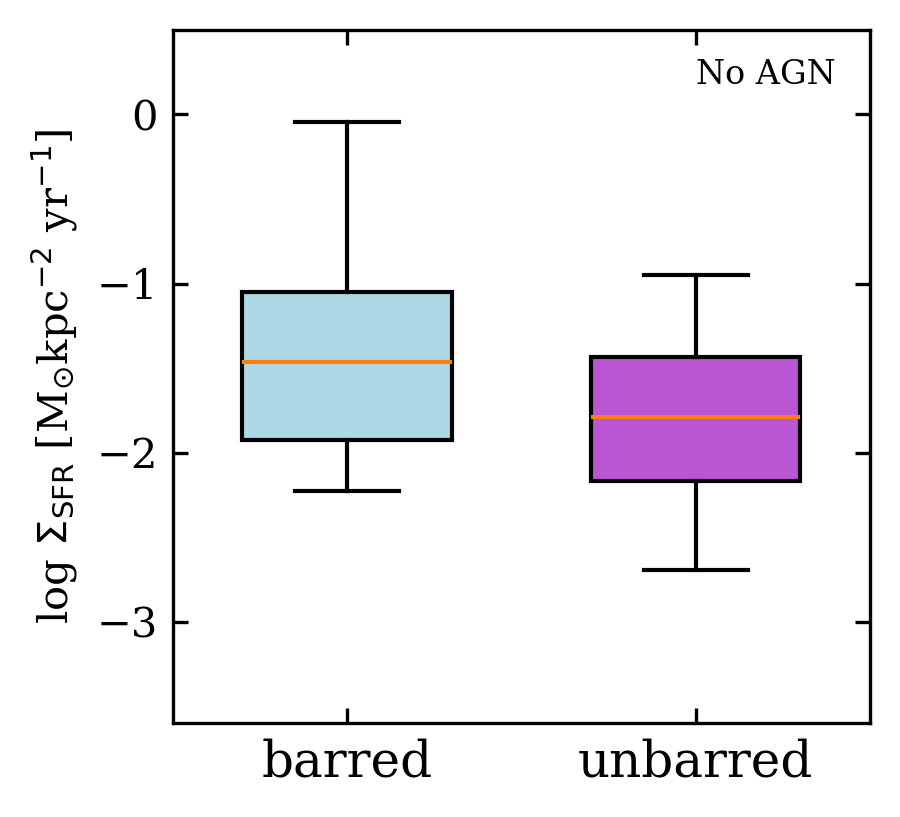}
    \includegraphics[height=0.17\textwidth,trim=0cm 0cm 0cm 0cm,clip]{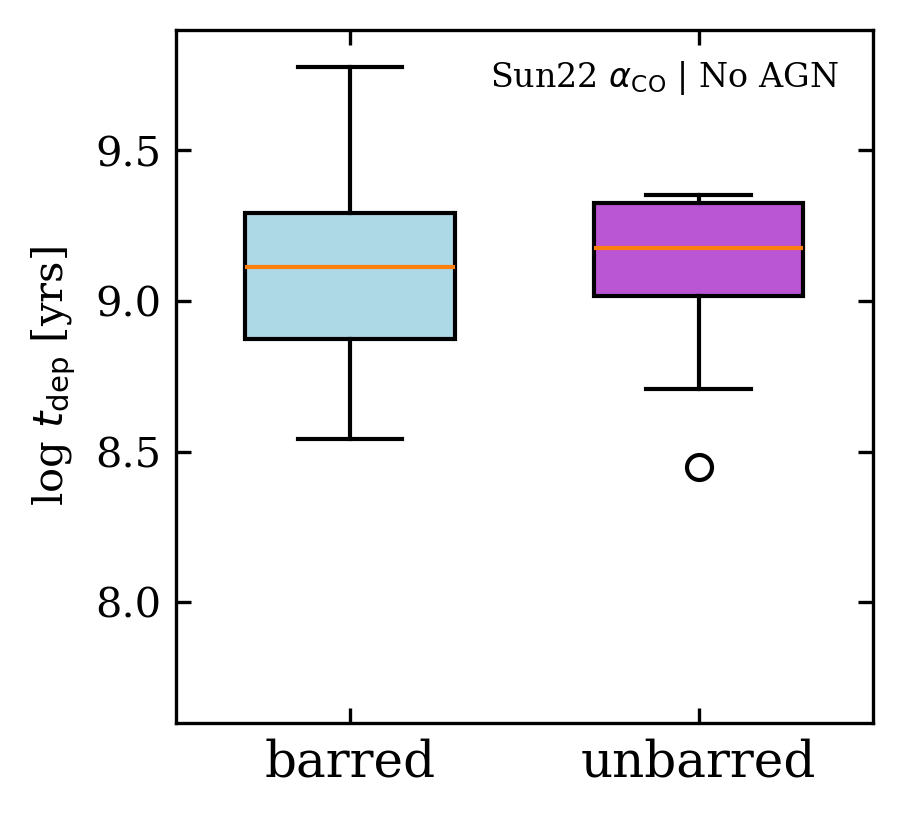}

    \caption{Boxplots comparing the distributions of the full sample of barred (42) and unbarred (22) galaxies [top row] and sample with no AGN [bottom row] for each property, using the metallicity dependent $\alpha_{\rm{CO}}$. See Figure \ref{fig:boxResolvedT24} for description of the quantities and method. The most significant differences between barred and unbarred galaxies for both the full sample and the sample with no AGN, are found for $\sigma_{\rm{v}}$ and $\Sigma_{\rm{mol,IW}}$ (See Table \ref{tab:ADwholeSample}). }
\label{fig:boxResolvedMetals}
\end{figure*}

Figure \ref{fig:boxResolvedMetals} shows each of the five properties in boxplots comparing the barred and unbarred galaxy distributions with the full sample in the top row and the subset with no AGN in the bottom row. In the full sample, we see a significantly higher median value for barred galaxies compared to unbarred galaxies for the first four properties, where this is most prominent for the intensity-weighted $\Sigma_{\rm{mol,IW}}$ and $\sigma_{\rm{v}}$. In contrast to Figure \ref{fig:boxResolvedT24}, there is no difference between the distributions for the fifth boxplot showing $t_{\rm dep}$. The patterns are similar for the sample of galaxies once AGN are removed. 

As in the main text, these trends are confirmed in the results of the AD-tests comparing barred and unbarred galaxies, shown in Table \ref{tab:ADwholeSample}. For the full sample, the first four properties ($\Sigma_{\rm{mol,RA}}$, $\Sigma_{\rm{mol,IW}}$, $\sigma_{\rm{v}}$, and $\Sigma_{\rm{SFR}}$) show moderate to strong evidence against the null hypothesis suggesting that they are not from the same population. In this case, the AD-test for $t_{\rm dep}$ and $\Sigma_{*}$ suggests that the central pixels of the barred and unbarred populations are the same.

\begin{deluxetable}{lcccc}
\tablenum{B1}
\tablecaption{Comparing barred and unbarred galaxy centres: full sample and sample which excludes galaxies with an AGN using metallicity-based $\alpha_{\rm{CO}}$ from \cite{SunJiayi2022multi}. \label{tab:ADwholeSample}}
\tablewidth{0pt}
\tablehead{
\colhead{} & \multicolumn{2}{c}{Full Sample} & \multicolumn{2}{c}{No AGN} \\
\multicolumn{1}{l}{Property} & \colhead{AD stat} & \colhead{$p$-value} & \colhead{AD stat} & \colhead{$p$-value} 
} 
\startdata
$\Sigma_{\rm{mol,RA}}$ & 3.6 & 0.0115 & 1.3 & 0.09    \\ 
$\Sigma_{\rm{mol,IW}}$ & 6.8 & 0.00078 & 4.0 & 0.0087 \\
$\sigma_{\rm{v}}$ & 8.3 & 0.00023 & 5.4 & 0.0025  \\
$\Sigma_{\rm{SFR}}$ & 3.5 & 0.013 & 1.8 & 0.058  \\ 
$\Sigma_{*}$ & 1.7 & 0.065 & 0.38 & 0.24  \\
$t_{\rm{dep}}$ & 0.034 & 0.34 & -0.15 & 0.42   \\
$\sigma_{\rm{v}}$/$\Sigma_{\rm{mol,IW}}$ & 4.2 & 0.00703 & 1.9 & 0.051  \\ 
$\sigma_{\rm{v}}$/$\Sigma_{\rm{SFR}}$ & 0.12 & 0.31 & -0.49 & 0.62 \\
\enddata
\tablecomments{Statistical results of the Anderson-Darling test comparing 42 barred, 22 unbarred galaxies (full sample) and 30 barred, 19 unbarred galaxies with no AGN for each property. We show the AD statistic and $p$-value. Quantities with $p$-values $<0.05$ have evidence against the null hypothesis, which implies that barred and unbarred galaxy centres are likely significantly different populations. In the comparison for galaxies with no AGN only $\Sigma_{\rm{mol,IW}}$ and the $\sigma_{\rm{v}}$ have $p$-values $<0.05$ implying barred and unbarred galaxy centres are likely significantly different populations. }
\end{deluxetable}

\begin{figure}
	\centering
    \includegraphics[width=0.45\textwidth,trim=0cm 0cm 0cm 1.0cm,clip]{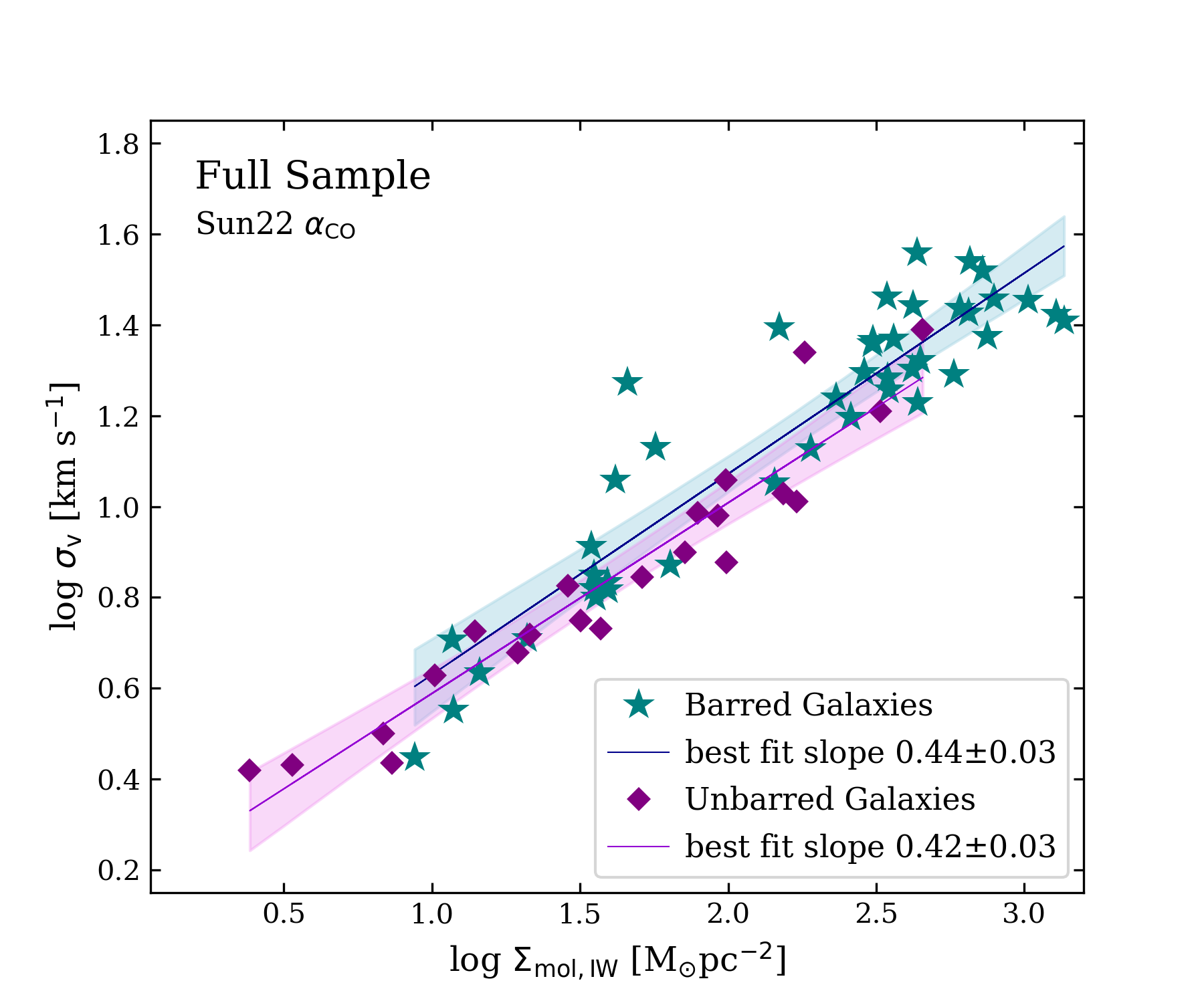} \\
    \includegraphics[width=0.45\textwidth]{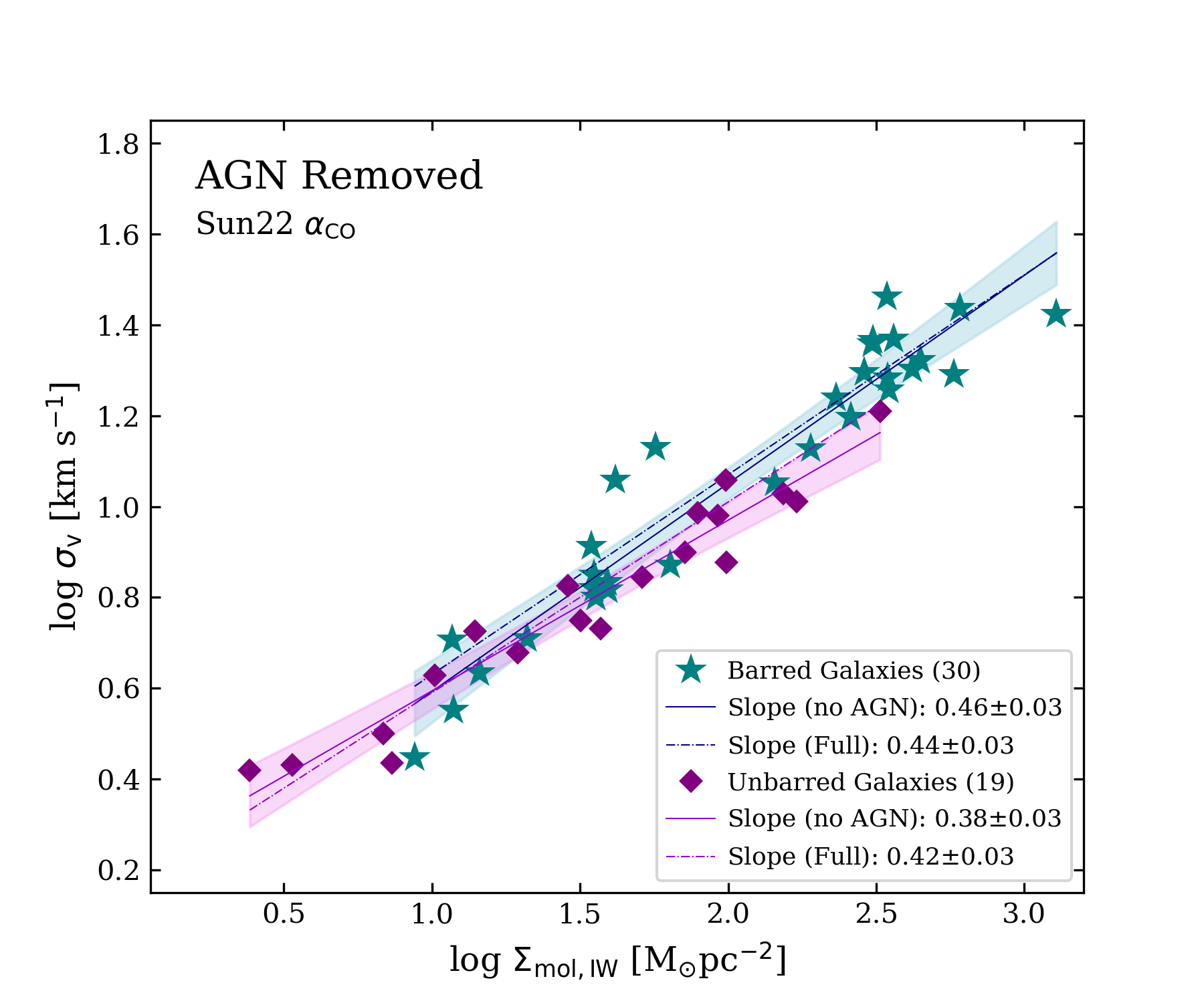}
    \caption{Barred versus unbarred resolved galaxy properties in the central region using the metallicity-based $\alpha_{\rm{CO}}$ from \cite{SunJiayi2022multi} [Top] $\sigma_{\rm{v}}$ as a function of $\Sigma_{\rm{mol,IW}}$ for the full sample; [Bottom] $\sigma_{\rm{v}}$ as a function of $\Sigma_{\rm{mol,IW}}$ for galaxies without an AGN. Best fit lines generated with Linmix with slope and intercept uncertainties are shown in shaded regions. Barred galaxies without AGN still show increased $\sigma_{\rm{v}}$ in the central region.}
    \label{fig:ResProp_Linmix_vdispSigmolRAIW}
\end{figure}

We perform a linear regression analysis using Linmix to generate the most probable best fit line for three different combinations of physical quantities ($\sigma_{\rm{v}}$ vs $\Sigma_{\rm{mol,IW}}$; $\Sigma_{\rm{SFR}}$ vs $\Sigma_{\rm{mol,RA}}$; and $\sigma_{\rm{v}}$ vs $\Sigma_{\rm{SFR}}$). Figure \ref{fig:ResProp_Linmix_vdispSigmolRAIW} shows the linear regression results for $\sigma_{\rm{v}}$ vs $\Sigma_{\rm{mol,IW}}$ for the full sample (upper) and the subset sample with no AGN (lower). 

In these plots, the fitted lines are very similar (compare to Figure \ref{fig:ResProp_Linmix_vdispSigmolIW_nAGN_T24}). 
However, we see the same higher values for barred galaxy centres as \cite{SunJiayi2020} who use the same metallicity-based $\alpha_{\rm{CO}}$.

\bibliography{BarsPaper}{}

\begin{thebibliography}{}
\expandafter\ifx\csname natexlab\endcsname\relax\def\natexlab#1{#1}\fi
\providecommand{\url}[1]{\href{#1}{#1}}
\providecommand{\dodoi}[1]{doi:~\href{http://doi.org/#1}{\nolinkurl{#1}}}
\providecommand{\doeprint}[1]{\href{http://ascl.net/#1}{\nolinkurl{http://ascl.net/#1}}}
\providecommand{\doarXiv}[1]{\href{https://arxiv.org/abs/#1}{\nolinkurl{https://arxiv.org/abs/#1}}}

\bibitem[{{Astropy Collaboration} {et~al.}(2013){Astropy Collaboration}, {Robitaille}, {Tollerud}, {Greenfield}, {Droettboom}, {Bray}, {Aldcroft}, {Davis}, {Ginsburg}, {Price-Whelan}, {Kerzendorf}, {Conley}, {Crighton}, {Barbary}, {Muna}, {Ferguson}, {Grollier}, {Parikh}, {Nair}, {Unther}, {Deil}, {Woillez}, {Conseil}, {Kramer}, {Turner}, {Singer}, {Fox}, {Weaver}, {Zabalza}, {Edwards}, {Azalee Bostroem}, {Burke}, {Casey}, {Crawford}, {Dencheva}, {Ely}, {Jenness}, {Labrie}, {Lim}, {Pierfederici}, {Pontzen}, {Ptak}, {Refsdal}, {Servillat}, \& {Streicher}}]{astropy2013}
{Astropy Collaboration}, {Robitaille}, T.~P., {Tollerud}, E.~J., {et~al.} 2013, \aap, 558, A33, \dodoi{10.1051/0004-6361/201322068}

\bibitem[{{Astropy Collaboration} {et~al.}(2018){Astropy Collaboration}, {Price-Whelan}, {Sip{\H{o}}cz}, {G{\"u}nther}, {Lim}, {Crawford}, {Conseil}, {Shupe}, {Craig}, {Dencheva}, {Ginsburg}, {VanderPlas}, {Bradley}, {P{\'e}rez-Su{\'a}rez}, {de Val-Borro}, {Aldcroft}, {Cruz}, {Robitaille}, {Tollerud}, {Ardelean}, {Babej}, {Bach}, {Bachetti}, {Bakanov}, {Bamford}, {Barentsen}, {Barmby}, {Baumbach}, {Berry}, {Biscani}, {Boquien}, {Bostroem}, {Bouma}, {Brammer}, {Bray}, {Breytenbach}, {Buddelmeijer}, {Burke}, {Calderone}, {Cano Rodr{\'\i}guez}, {Cara}, {Cardoso}, {Cheedella}, {Copin}, {Corrales}, {Crichton}, {D'Avella}, {Deil}, {Depagne}, {Dietrich}, {Donath}, {Droettboom}, {Earl}, {Erben}, {Fabbro}, {Ferreira}, {Finethy}, {Fox}, {Garrison}, {Gibbons}, {Goldstein}, {Gommers}, {Greco}, {Greenfield}, {Groener}, {Grollier}, {Hagen}, {Hirst}, {Homeier}, {Horton}, {Hosseinzadeh}, {Hu}, {Hunkeler}, {Ivezi{\'c}}, {Jain}, {Jenness}, {Kanarek}, {Kendrew}, {Kern}, {Kerzendorf}, {Khvalko}, {King}, {Kirkby}, {Kulkarni},
  {Kumar}, {Lee}, {Lenz}, {Littlefair}, {Ma}, {Macleod}, {Mastropietro}, {McCully}, {Montagnac}, {Morris}, {Mueller}, {Mumford}, {Muna}, {Murphy}, {Nelson}, {Nguyen}, {Ninan}, {N{\"o}the}, {Ogaz}, {Oh}, {Parejko}, {Parley}, {Pascual}, {Patil}, {Patil}, {Plunkett}, {Prochaska}, {Rastogi}, {Reddy Janga}, {Sabater}, {Sakurikar}, {Seifert}, {Sherbert}, {Sherwood-Taylor}, {Shih}, {Sick}, {Silbiger}, {Singanamalla}, {Singer}, {Sladen}, {Sooley}, {Sornarajah}, {Streicher}, {Teuben}, {Thomas}, {Tremblay}, {Turner}, {Terr{\'o}n}, {van Kerkwijk}, {de la Vega}, {Watkins}, {Weaver}, {Whitmore}, {Woillez}, {Zabalza}, \& {Astropy Contributors}}]{astropy2018}
{Astropy Collaboration}, {Price-Whelan}, A.~M., {Sip{\H{o}}cz}, B.~M., {et~al.} 2018, \aj, 156, 123, \dodoi{10.3847/1538-3881/aabc4f}

\bibitem[{{Astropy Collaboration} {et~al.}(2022){Astropy Collaboration}, {Price-Whelan}, {Lim}, {Earl}, {Starkman}, {Bradley}, {Shupe}, {Patil}, {Corrales}, {Brasseur}, {N{\"o}the}, {Donath}, {Tollerud}, {Morris}, {Ginsburg}, {Vaher}, {Weaver}, {Tocknell}, {Jamieson}, {van Kerkwijk}, {Robitaille}, {Merry}, {Bachetti}, {G{\"u}nther}, {Aldcroft}, {Alvarado-Montes}, {Archibald}, {B{\'o}di}, {Bapat}, {Barentsen}, {Baz{\'a}n}, {Biswas}, {Boquien}, {Burke}, {Cara}, {Cara}, {Conroy}, {Conseil}, {Craig}, {Cross}, {Cruz}, {D'Eugenio}, {Dencheva}, {Devillepoix}, {Dietrich}, {Eigenbrot}, {Erben}, {Ferreira}, {Foreman-Mackey}, {Fox}, {Freij}, {Garg}, {Geda}, {Glattly}, {Gondhalekar}, {Gordon}, {Grant}, {Greenfield}, {Groener}, {Guest}, {Gurovich}, {Handberg}, {Hart}, {Hatfield-Dodds}, {Homeier}, {Hosseinzadeh}, {Jenness}, {Jones}, {Joseph}, {Kalmbach}, {Karamehmetoglu}, {Ka{\l}uszy{\'n}ski}, {Kelley}, {Kern}, {Kerzendorf}, {Koch}, {Kulumani}, {Lee}, {Ly}, {Ma}, {MacBride}, {Maljaars}, {Muna}, {Murphy}, {Norman},
  {O'Steen}, {Oman}, {Pacifici}, {Pascual}, {Pascual-Granado}, {Patil}, {Perren}, {Pickering}, {Rastogi}, {Roulston}, {Ryan}, {Rykoff}, {Sabater}, {Sakurikar}, {Salgado}, {Sanghi}, {Saunders}, {Savchenko}, {Schwardt}, {Seifert-Eckert}, {Shih}, {Jain}, {Shukla}, {Sick}, {Simpson}, {Singanamalla}, {Singer}, {Singhal}, {Sinha}, {Sip{\H{o}}cz}, {Spitler}, {Stansby}, {Streicher}, {{\v{S}}umak}, {Swinbank}, {Taranu}, {Tewary}, {Tremblay}, {de Val-Borro}, {Van Kooten}, {Vasovi{\'c}}, {Verma}, {de Miranda Cardoso}, {Williams}, {Wilson}, {Winkel}, {Wood-Vasey}, {Xue}, {Yoachim}, {Zhang}, {Zonca}, \& {Astropy Project Contributors}}]{astropy2022}
{Astropy Collaboration}, {Price-Whelan}, A.~M., {Lim}, P.~L., {et~al.} 2022, \apj, 935, 167, \dodoi{10.3847/1538-4357/ac7c74}

\bibitem[{{Bolatto} {et~al.}(2008){Bolatto}, {Leroy}, {Rosolowsky}, {Walter}, \& {Blitz}}]{Bolatto2008}
{Bolatto}, A.~D., {Leroy}, A.~K., {Rosolowsky}, E., {Walter}, F., \& {Blitz}, L. 2008, ApJ, 686, 948, \dodoi{10.1086/591513}

\bibitem[{{Bolatto} {et~al.}(2013){Bolatto}, {Wolfire}, \& {Leroy}}]{Bolatto2013_KeyCO}
{Bolatto}, A.~D., {Wolfire}, M., \& {Leroy}, A.~K. 2013, ARA\&A, 51, 207, \dodoi{10.1146/annurev-astro-082812-140944}

\bibitem[{{Bolatto} {et~al.}(2021){Bolatto}, {Leroy}, {Levy}, {Meier}, {Mills}, {Thompson}, {Emig}, {Veilleux}, {Ott}, {Gorski}, {Walter}, {Lopez}, \& {Lenki{\'c}}}]{Bolatto2021}
{Bolatto}, A.~D., {Leroy}, A.~K., {Levy}, R.~C., {et~al.} 2021, ApJ, 923, 83, \dodoi{10.3847/1538-4357/ac2c08}

\bibitem[{{Brown} {et~al.}(2023){Brown}, {Roberts}, {Thorp}, {Ellison}, {Zabel}, {Wilson}, {Bah{\'e}}, {Bisaria}, {Bolatto}, {Boselli}, {Chung}, {Cortese}, {Catinella}, {Davis}, {Jim{\'e}nez-Donaire}, {Lagos}, {Lee}, {Parker}, {Smith}, {Spekkens}, {Stevens}, {Villanueva}, \& {Watts}}]{Brown2023}
{Brown}, T., {Roberts}, I.~D., {Thorp}, M., {et~al.} 2023, ApJ, 956, 37, \dodoi{10.3847/1538-4357/acf195}

\bibitem[{{Chevance} {et~al.}(2023){Chevance}, {Krumholz}, {McLeod}, {Ostriker}, {Rosolowsky}, \& {Sternberg}}]{Chevance2023}
{Chevance}, M., {Krumholz}, M.~R., {McLeod}, A.~F., {et~al.} 2023, in Astronomical Society of the Pacific Conference Series, Vol. 534, Protostars and Planets VII, ed. S.~{Inutsuka}, Y.~{Aikawa}, T.~{Muto}, K.~{Tomida}, \& M.~{Tamura}, 1, \dodoi{10.48550/arXiv.2203.09570}

\bibitem[{{Choi} {et~al.}(2023){Choi}, {Liu}, {Bureau}, {Cappellari}, {Davis}, {Gensior}, {Liang}, {Lu}, {Williams}, \& {Chung}}]{Choi2023}
{Choi}, W., {Liu}, L., {Bureau}, M., {et~al.} 2023, MNRAS, 522, 4078, \dodoi{10.1093/mnras/stad1211}

\bibitem[{{Combes} {et~al.}(2019){Combes}, {Garc{\'\i}a-Burillo}, {Audibert}, {Hunt}, {Eckart}, {Aalto}, {Casasola}, {Boone}, {Krips}, {Viti}, {Sakamoto}, {Muller}, {Dasyra}, {van der Werf}, \& {Martin}}]{Combes2019}
{Combes}, F., {Garc{\'\i}a-Burillo}, S., {Audibert}, A., {et~al.} 2019, A\&A, 623, A79, \dodoi{10.1051/0004-6361/201834560}

\bibitem[{{Costa} {et~al.}(2014){Costa}, {Sijacki}, \& {Haehnelt}}]{Costa2014}
{Costa}, T., {Sijacki}, D., \& {Haehnelt}, M.~G. 2014, MNRAS, 444, 2355, \dodoi{10.1093/mnras/stu1632}

\bibitem[{{de Vaucouleurs}(1963)}]{deVaucouleurs1963}
{de Vaucouleurs}, G. 1963, ApJS, 8, 31, \dodoi{10.1086/190084}

\bibitem[{{D{\'\i}az-Garc{\'\i}a} {et~al.}(2020){D{\'\i}az-Garc{\'\i}a}, {Moyano}, {Comer{\'o}n}, {Knapen}, {Salo}, \& {Bouquin}}]{Diaz-Garcia2020distSF}
{D{\'\i}az-Garc{\'\i}a}, S., {Moyano}, F.~D., {Comer{\'o}n}, S., {et~al.} 2020, A\&A, 644, A38, \dodoi{10.1051/0004-6361/202039162}

\bibitem[{{D{\'\i}az-Garc{\'\i}a} {et~al.}(2021){D{\'\i}az-Garc{\'\i}a}, {Lisenfeld}, {P{\'e}rez}, {Zurita}, {Verley}, {Combes}, {Espada}, {Leon}, {Mart{\'\i}nez-Badenes}, {Sabater}, \& {Verdes-Montenegro}}]{Diaz-Garcia2021}
{D{\'\i}az-Garc{\'\i}a}, S., {Lisenfeld}, U., {P{\'e}rez}, I., {et~al.} 2021, A\&A, 654, A135, \dodoi{10.1051/0004-6361/202140674}

\bibitem[{{Garland} {et~al.}(2024){Garland}, {Walmsley}, {Silcock}, {Potts}, {Smith}, {Simmons}, {Lintott}, {Smethurst}, {Dawson}, {Keel}, {Kruk}, {Mantha}, {Masters}, {O'Ryan}, {Popp}, \& {Thorne}}]{Garland2024}
{Garland}, I.~L., {Walmsley}, M., {Silcock}, M.~S., {et~al.} 2024, MNRAS, \dodoi{10.1093/mnras/stae1620}

\bibitem[{{Ginsburg} {et~al.}(2019){Ginsburg}, {Koch}, {Robitaille}, {Beaumont}, {Adamginsburg}, {Sip{\H{o}}cz}, {ZuHone}, {Patra}, {Jones}, {Lim}, {Stern}, {Rosolowsky}, {Earl}, {De Val-Borro}, {Jrobbfed}, {Shuokong}, {Kepley}, {Sokolov}, {Badger}, {Maret}, {Garrido}, {Booker}, \& {Tollerud}}]{Ginsburg2019}
{Ginsburg}, A., {Koch}, E., {Robitaille}, T., {et~al.} 2019, {radio-astro-tools/spectral-cube: Release v0.4.5}, v0.4.5, Zenodo,  Zenodo, \dodoi{10.5281/zenodo.3558614}

\bibitem[{Harris {et~al.}(2020)Harris, Millman, van~der Walt, Gommers, Virtanen, Cournapeau, Wieser, Taylor, Berg, Smith, Kern, Picus, Hoyer, van Kerkwijk, Brett, Haldane, del R{\'{i}}o, Wiebe, Peterson, G{\'{e}}rard-Marchant, Sheppard, Reddy, Weckesser, Abbasi, Gohlke, \& Oliphant}]{harris2020}
Harris, C.~R., Millman, K.~J., van~der Walt, S.~J., {et~al.} 2020, Nature, 585, 357, \dodoi{10.1038/s41586-020-2649-2}

\bibitem[{{Heyer} {et~al.}(2001){Heyer}, {Carpenter}, \& {Snell}}]{Heyer2001ApJ}
{Heyer}, M.~H., {Carpenter}, J.~M., \& {Snell}, R.~L. 2001, ApJ, 551, 852, \dodoi{10.1086/320218}

\bibitem[{{Hogarth} {et~al.}(2024){Hogarth}, {Saintonge}, {Davis}, {Ellison}, {Lin}, {L{\'o}pez-Cob{\'a}}, {Pan}, \& {Thorp}}]{Hogarth2024}
{Hogarth}, L.~M., {Saintonge}, A., {Davis}, T.~A., {et~al.} 2024, MNRAS, 528, 6768, \dodoi{10.1093/mnras/stae377}

\bibitem[{Hunter(2007)}]{Hunter2007}
Hunter, J.~D. 2007, Computing in Science \& Engineering, 9, 90, \dodoi{10.1109/MCSE.2007.55}

\bibitem[{{Israel}(2020)}]{Israel2020}
{Israel}, F.~P. 2020, A\&A, 635, A131, \dodoi{10.1051/0004-6361/201834198}

\bibitem[{{Jim{\'e}nez-Donaire} {et~al.}(2023){Jim{\'e}nez-Donaire}, {Brown}, {Wilson}, {Roberts}, {Zabel}, {Ellison}, {Thorp}, {Villanueva}, {Chown}, {Bisaria}, {Bolatto}, {Boselli}, {Catinella}, {Chung}, {Cortese}, {Davis}, {Lagos}, {Lee}, {Parker}, {Spekkens}, {Stevens}, \& {Sun}}]{Jimenez-Donaire2023}
{Jim{\'e}nez-Donaire}, M.~J., {Brown}, T., {Wilson}, C.~D., {et~al.} 2023, A\&A, 671, A3, \dodoi{10.1051/0004-6361/202244718}

\bibitem[{{Jogee} {et~al.}(2005){Jogee}, {Scoville}, \& {Kenney}}]{Jogee2005-Cntr}
{Jogee}, S., {Scoville}, N., \& {Kenney}, J. D.~P. 2005, ApJL, 630, 837, \dodoi{10.1086/432106}

\bibitem[{{Jogee} {et~al.}(2004){Jogee}, {Barazza}, {Rix}, {Shlosman}, {Barden}, {Wolf}, {Davies}, {Heyer}, {Beckwith}, {Bell}, {Borch}, {Caldwell}, {Conselice}, {Dahlen}, {H{\"a}ussler}, {Heymans}, {Jahnke}, {Knapen}, {Laine}, {Lubell}, {Mobasher}, {McIntosh}, {Meisenheimer}, {Peng}, {Ravindranath}, {Sanchez}, {Somerville}, \& {Wisotzki}}]{Jogee2004-8Gyr}
{Jogee}, S., {Barazza}, F.~D., {Rix}, H.-W., {et~al.} 2004, ApJL, 615, L105, \dodoi{10.1086/426138}

\bibitem[{{Kormendy} \& {Kennicutt}(2004)}]{Kormendy2004secularEvol}
{Kormendy}, J., \& {Kennicutt}, Robert~C., J. 2004, ARA\&A, 42, 603, \dodoi{10.1146/annurev.astro.42.053102.134024}

\bibitem[{{Krumholz}(2021)}]{Krumholz2021}
{Krumholz}, M.~R. 2021, in Astronomical Society of the Pacific Conference Series, Vol. 528, New Horizons in Galactic Center Astronomy and Beyond, ed. M.~{Tsuboi} \& T.~{Oka}, 19

\bibitem[{{Leroy} {et~al.}(2019){Leroy}, {Sandstrom}, {Lang}, {Lewis}, {Salim}, {Behrens}, {Chastenet}, {Chiang}, {Gallagher}, {Kessler}, \& {Utomo}}]{Leroy2019}
{Leroy}, A.~K., {Sandstrom}, K.~M., {Lang}, D., {et~al.} 2019, ApJS, 244, 24, \dodoi{10.3847/1538-4365/ab3925}

\bibitem[{{Leroy} {et~al.}(2021){Leroy}, {Schinnerer}, {Hughes}, {Rosolowsky}, {Pety}, {Schruba}, {Usero}, {Blanc}, {Chevance}, {Emsellem}, {Faesi}, {Herrera}, {Liu}, {Meidt}, {Querejeta}, {Saito}, {Sandstrom}, {Sun}, {Williams}, {Anand}, {Barnes}, {Behrens}, {Belfiore}, {Benincasa}, {Be{\v{s}}li{\'c}}, {Bigiel}, {Bolatto}, {den Brok}, {Cao}, {Chandar}, {Chastenet}, {Chiang}, {Congiu}, {Dale}, {Deger}, {Eibensteiner}, {Egorov}, {Garc{\'\i}a-Rodr{\'\i}guez}, {Glover}, {Grasha}, {Henshaw}, {Ho}, {Kepley}, {Kim}, {Klessen}, {Kreckel}, {Koch}, {Kruijssen}, {Larson}, {Lee}, {Lopez}, {Machado}, {Mayker}, {McElroy}, {Murphy}, {Ostriker}, {Pan}, {Pessa}, {Puschnig}, {Razza}, {S{\'a}nchez-Bl{\'a}zquez}, {Santoro}, {Sardone}, {Scheuermann}, {Sliwa}, {Sormani}, {Stuber}, {Thilker}, {Turner}, {Utomo}, {Watkins}, \& {Whitmore}}]{Leroy2021b}
{Leroy}, A.~K., {Schinnerer}, E., {Hughes}, A., {et~al.} 2021, ApJS, 257, 43, \dodoi{10.3847/1538-4365/ac17f3}

\bibitem[{{Liu} {et~al.}(2023){Liu}, {Schinnerer}, {Saito}, {Rosolowsky}, {Leroy}, {Usero}, {Sandstrom}, {Klessen}, {Glover}, {Ao}, {Be{\v{s}}li{\'c}}, {Bigiel}, {Cao}, {Chastenet}, {Chevance}, {Dale}, {Gao}, {Hughes}, {Kreckel}, {Kruijssen}, {Pan}, {Pety}, {Salak}, {Santoro}, {Schruba}, {Sun}, {Teng}, \& {Williams}}]{Liu2023}
{Liu}, D., {Schinnerer}, E., {Saito}, T., {et~al.} 2023, A\&A, 672, A36, \dodoi{10.1051/0004-6361/202244564}

\bibitem[{{Liu} {et~al.}(2021){Liu}, {Bureau}, {Blitz}, {Davis}, {Onishi}, {Smith}, {North}, \& {Iguchi}}]{Liu2021}
{Liu}, L., {Bureau}, M., {Blitz}, L., {et~al.} 2021, \mnras, 505, 4048, \dodoi{10.1093/mnras/stab1537}

\bibitem[{{Maeda} {et~al.}(2023){Maeda}, {Egusa}, {Ohta}, {Fujimoto}, \& {Habe}}]{Maeda2023SFgasrichBars}
{Maeda}, F., {Egusa}, F., {Ohta}, K., {Fujimoto}, Y., \& {Habe}, A. 2023, ApJ, 943, 7, \dodoi{10.3847/1538-4357/aca664}

\bibitem[{{McMullin} {et~al.}(2007){McMullin}, {Waters}, {Schiebel}, {Young}, \& {Golap}}]{McMullin2007}
{McMullin}, J.~P., {Waters}, B., {Schiebel}, D., {Young}, W., \& {Golap}, K. 2007, in Astronomical Society of the Pacific Conference Series, Vol. 376, Astronomical Data Analysis Software and Systems XVI, ed. R.~A. {Shaw}, F.~{Hill}, \& D.~J. {Bell}, 127

\bibitem[{{Miville-Desch{\^e}nes} {et~al.}(2017){Miville-Desch{\^e}nes}, {Murray}, \& {Lee}}]{2017MivilleDeschenes}
{Miville-Desch{\^e}nes}, M.-A., {Murray}, N., \& {Lee}, E.~J. 2017, \apj, 834, 57, \dodoi{10.3847/1538-4357/834/1/57}

\bibitem[{{North} {et~al.}(2019){North}, {Davis}, {Bureau}, {Cappellari}, {Iguchi}, {Liu}, {Onishi}, {Sarzi}, {Smith}, \& {Williams}}]{North2019}
{North}, E.~V., {Davis}, T.~A., {Bureau}, M., {et~al.} 2019, MNRAS, 490, 319, \dodoi{10.1093/mnras/stz2598}

\bibitem[{{Querejeta} {et~al.}(2021){Querejeta}, {Schinnerer}, {Meidt}, {Sun}, {Leroy}, {Emsellem}, {Klessen}, {Mu{\~n}oz-Mateos}, {Salo}, {Laurikainen}, {Be{\v{s}}li{\'c}}, {Blanc}, {Chevance}, {Dale}, {Eibensteiner}, {Faesi}, {Garc{\'\i}a-Rodr{\'\i}guez}, {Glover}, {Grasha}, {Henshaw}, {Herrera}, {Hughes}, {Kreckel}, {Kruijssen}, {Liu}, {Murphy}, {Pan}, {Pety}, {Razza}, {Rosolowsky}, {Saito}, {Schruba}, {Usero}, {Watkins}, \& {Williams}}]{Querejeta2021}
{Querejeta}, M., {Schinnerer}, E., {Meidt}, S., {et~al.} 2021, A\&A, 656, A133, \dodoi{10.1051/0004-6361/202140695}

\bibitem[{{R Core Team}(2023)}]{R2023}
{R Core Team}. 2023, R: A Language and Environment for Statistical Computing, R Foundation for Statistical Computing, Vienna, Austria.
\newblock \url{https://www.R-project.org/}

\bibitem[{{Rosolowsky} {et~al.}(2021){Rosolowsky}, {Hughes}, {Leroy}, {Sun}, {Querejeta}, {Schruba}, {Usero}, {Herrera}, {Liu}, {Pety}, {Saito}, {Be{\v{s}}li{\'c}}, {Bigiel}, {Blanc}, {Chevance}, {Dale}, {Deger}, {Faesi}, {Glover}, {Henshaw}, {Klessen}, {Kruijssen}, {Larson}, {Lee}, {Meidt}, {Mok}, {Schinnerer}, {Thilker}, \& {Williams}}]{Rosolowsky2021MNRAS}
{Rosolowsky}, E., {Hughes}, A., {Leroy}, A.~K., {et~al.} 2021, MNRAS, 502, 1218, \dodoi{10.1093/mnras/stab085}

\bibitem[{{Sakamoto} {et~al.}(1999){Sakamoto}, {Okumura}, {Ishizuki}, \& {Scoville}}]{Sakamoto1999}
{Sakamoto}, K., {Okumura}, S.~K., {Ishizuki}, S., \& {Scoville}, N.~Z. 1999, ApJ, 525, 691, \dodoi{10.1086/307910}

\bibitem[{{Sandstrom} {et~al.}(2013){Sandstrom}, {Leroy}, {Walter}, {Bolatto}, {Croxall}, {Draine}, {Wilson}, {Wolfire}, {Calzetti}, {Kennicutt}, {Aniano}, {Donovan Meyer}, {Usero}, {Bigiel}, {Brinks}, {de Blok}, {Crocker}, {Dale}, {Engelbracht}, {Galametz}, {Groves}, {Hunt}, {Koda}, {Kreckel}, {Linz}, {Meidt}, {Pellegrini}, {Rix}, {Roussel}, {Schinnerer}, {Schruba}, {Schuster}, {Skibba}, {van der Laan}, {Appleton}, {Armus}, {Brandl}, {Gordon}, {Hinz}, {Krause}, {Montiel}, {Sauvage}, {Schmiedeke}, {Smith}, \& {Vigroux}}]{Sandstrom2013}
{Sandstrom}, K.~M., {Leroy}, A.~K., {Walter}, F., {et~al.} 2013, ApJ, 777, 5, \dodoi{10.1088/0004-637X/777/1/5}

\bibitem[{{Schinnerer} {et~al.}(2019){Schinnerer}, {Hughes}, {Leroy}, {Groves}, {Blanc}, {Kreckel}, {Bigiel}, {Chevance}, {Dale}, {Emsellem}, {Faesi}, {Glover}, {Grasha}, {Henshaw}, {Hygate}, {Kruijssen}, {Meidt}, {Pety}, {Querejeta}, {Rosolowsky}, {Saito}, {Schruba}, {Sun}, \& {Utomo}}]{Schinnerer2019}
{Schinnerer}, E., {Hughes}, A., {Leroy}, A., {et~al.} 2019, ApJ, 887, 49, \dodoi{10.3847/1538-4357/ab50c2}

\bibitem[{{Scholz} \& {Stephens}(1987)}]{Scholz1987Stats}
{Scholz}, F.~W., \& {Stephens}, M.~A. 1987, Journal of the American Statistical Association, 82, 918, \dodoi{10.1080/01621459.1987.10478517}

\bibitem[{{Sheth} {et~al.}(2002){Sheth}, {Vogel}, {Teuben}, {Harris}, {Regan}, {Thornley}, \& {Helfer}}]{Sheth2002CompMolGas}
{Sheth}, K., {Vogel}, S.~N., {Teuben}, P.~J., {et~al.} 2002, in Astronomical Society of the Pacific Conference Series, Vol. 275, Disks of Galaxies: Kinematics, Dynamics and Peturbations, ed. E.~{Athanassoula}, A.~{Bosma}, \& R.~{Mujica}, 263--266

\bibitem[{{Sheth} {et~al.}(2010){Sheth}, {Regan}, {Hinz}, {Gil de Paz}, {Men{\'e}ndez-Delmestre}, {Mu{\~n}oz-Mateos}, {Seibert}, {Kim}, {Laurikainen}, {Salo}, {Gadotti}, {Laine}, {Mizusawa}, {Armus}, {Athanassoula}, {Bosma}, {Buta}, {Capak}, {Jarrett}, {Elmegreen}, {Elmegreen}, {Knapen}, {Koda}, {Helou}, {Ho}, {Madore}, {Masters}, {Mobasher}, {Ogle}, {Peng}, {Schinnerer}, {Surace}, {Zaritsky}, {Comer{\'o}n}, {de Swardt}, {Meidt}, {Kasliwal}, \& {Aravena}}]{Sheth2010}
{Sheth}, K., {Regan}, M., {Hinz}, J.~L., {et~al.} 2010, PASP, 122, 1397, \dodoi{10.1086/657638}

\bibitem[{{Stuber} {et~al.}(2021){Stuber}, {Saito}, {Schinnerer}, {Emsellem}, {Querejeta}, {Williams}, {Barnes}, {Bigiel}, {Blanc}, {Dale}, {Grasha}, {Klessen}, {Kruijssen}, {Leroy}, {Meidt}, {Pan}, {Rosolowsky}, {Schruba}, {Sun}, \& {Usero}}]{Stuber2021}
{Stuber}, S.~K., {Saito}, T., {Schinnerer}, E., {et~al.} 2021, A\&A, 653, A172, \dodoi{10.1051/0004-6361/202141093}

\bibitem[{{Sun} {et~al.}(2018){Sun}, {Leroy}, {Schruba}, {Rosolowsky}, {Hughes}, {Kruijssen}, {Meidt}, {Schinnerer}, {Blanc}, {Bigiel}, {Bolatto}, {Chevance}, {Groves}, {Herrera}, {Hygate}, {Pety}, {Querejeta}, {Usero}, \& {Utomo}}]{SunJiayi2018}
{Sun}, J., {Leroy}, A.~K., {Schruba}, A., {et~al.} 2018, ApJ, 860, 172, \dodoi{10.3847/1538-4357/aac326}

\bibitem[{{Sun} {et~al.}(2020){Sun}, {Leroy}, {Schinnerer}, {Hughes}, {Rosolowsky}, {Querejeta}, {Schruba}, {Liu}, {Saito}, {Herrera}, {Faesi}, {Usero}, {Pety}, {Kruijssen}, {Ostriker}, {Bigiel}, {Blanc}, {Bolatto}, {Boquien}, {Chevance}, {Dale}, {Deger}, {Emsellem}, {Glover}, {Grasha}, {Groves}, {Henshaw}, {Jimenez-Donaire}, {Kim}, {Klessen}, {Kreckel}, {Lee}, {Meidt}, {Sandstrom}, {Sardone}, {Utomo}, \& {Williams}}]{SunJiayi2020}
{Sun}, J., {Leroy}, A.~K., {Schinnerer}, E., {et~al.} 2020, ApJL, 901, L8, \dodoi{10.3847/2041-8213/abb3be}

\bibitem[{{Sun} {et~al.}(2022){Sun}, {Leroy}, {Rosolowsky}, {Hughes}, {Schinnerer}, {Schruba}, {Koch}, {Blanc}, {Chiang}, {Groves}, {Liu}, {Meidt}, {Pan}, {Pety}, {Querejeta}, {Saito}, {Sandstrom}, {Sardone}, {Usero}, {Utomo}, {Williams}, {Barnes}, {Benincasa}, {Bigiel}, {Bolatto}, {Boquien}, {Chevance}, {Dale}, {Deger}, {Emsellem}, {Glover}, {Grasha}, {Henshaw}, {Klessen}, {Kreckel}, {Kruijssen}, {Ostriker}, \& {Thilker}}]{SunJiayi2022multi}
{Sun}, J., {Leroy}, A.~K., {Rosolowsky}, E., {et~al.} 2022, AJ, 164, 43, \dodoi{10.3847/1538-3881/ac74bd}

\bibitem[{{Teng} {et~al.}(2023){Teng}, {Sandstrom}, {Sun}, {Gong}, {Bolatto}, {Chiang}, {Leroy}, {Usero}, {Glover}, {Klessen}, {Liu}, {Querejeta}, {Schinnerer}, {Bigiel}, {Cao}, {Chevance}, {Eibensteiner}, {Grasha}, {Israel}, {Murphy}, {Neumann}, {Pan}, {Pinna}, {Sormani}, {Smith}, {Walter}, \& {Williams}}]{Teng2023}
{Teng}, Y.-H., {Sandstrom}, K.~M., {Sun}, J., {et~al.} 2023, ApJ, 950, 119, \dodoi{10.3847/1538-4357/accb86}

\bibitem[{{Teng} {et~al.}(2024){Teng}, {Chiang}, {Sandstrom}, {Sun}, {Leroy}, {Bolatto}, {Usero}, {Ostriker}, {Querejeta}, {Chastenet}, {Bigiel}, {Boquien}, {den Brok}, {Cao}, {Chevance}, {Chown}, {Colombo}, {Eibensteiner}, {Glover}, {Grasha}, {Henshaw}, {Jim{\'e}nez-Donaire}, {Liu}, {Murphy}, {Pan}, {Stuber}, \& {Williams}}]{Teng2024}
{Teng}, Y.-H., {Chiang}, I.-D., {Sandstrom}, K.~M., {et~al.} 2024, ApJ, 961, 42, \dodoi{10.3847/1538-4357/ad10ae}

\bibitem[{{V{\'e}ron-Cetty} \& {V{\'e}ron}(2010)}]{Veron-Cetty2010}
{V{\'e}ron-Cetty}, M.~P., \& {V{\'e}ron}, P. 2010, A\&A, 518, A10, \dodoi{10.1051/0004-6361/201014188}

\bibitem[{Virtanen {et~al.}(2020)Virtanen, Gommers, Oliphant, Haberland, Reddy, Cournapeau, Burovski, Peterson, Weckesser, Bright, {van der Walt}, Brett, Wilson, Millman, Mayorov, Nelson, Jones, Kern, Larson, Carey, Polat, Feng, Moore, {VanderPlas}, Laxalde, Perktold, Cimrman, Henriksen, Quintero, Harris, Archibald, Ribeiro, Pedregosa, {van Mulbregt}, \& {SciPy 1.0 Contributors}}]{scipy2020}
Virtanen, P., Gommers, R., Oliphant, T.~E., {et~al.} 2020, Nature Methods, 17, 261, \dodoi{10.1038/s41592-019-0686-2}

\bibitem[{{Wilson} {et~al.}(2011){Wilson}, {Warren}, {Irwin}, {Knapen}, {Israel}, {Serjeant}, {Attewell}, {Bendo}, {Brinks}, {Butner}, {Clements}, {Leech}, {Matthews}, {M{\"u}hle}, {Mortier}, {Parkin}, {Petitpas}, {Tan}, {Tilanus}, {Usero}, {Vaccari}, {van der Werf}, {Wiegert}, \& {Zhu}}]{Wilson2011JCMT}
{Wilson}, C.~D., {Warren}, B.~E., {Irwin}, J., {et~al.} 2011, MNRAS, 410, 1409, \dodoi{10.1111/j.1365-2966.2010.17646.x}

\end{thebibliography}
\bibliographystyle{aasjournal}



\end{document}